\documentclass[10pt]{article}
\usepackage[utf8]{inputenc}
\usepackage[T1]{fontenc}
\usepackage{hyperref}       
\usepackage{url}            
\usepackage{booktabs}       
\usepackage{amsfonts}       
\usepackage{nicefrac}       
\usepackage{microtype}      
\usepackage{lipsum}
\usepackage{graphicx} 
\usepackage[short]{optidef}
\usepackage{bbold}
\usepackage{verbatim}
\usepackage{tikz}
\usetikzlibrary{shapes,arrows,shadows,decorations.pathreplacing,angles,quotes}  
\usepackage{subfigure}
\usepackage[left=1in,right=1in]{geometry}

\usepackage{amsmath}

\usetikzlibrary{shapes,arrows,shadows,decorations.pathreplacing,angles,quotes}

\usepackage{graphicx}
\usepackage{dcolumn}
\usepackage{bm}
\usepackage{authblk}
\begin{document}

\title{Filtering time-dependent covariance matrices using time-independent eigenvalues}

\author[1,*]{Christian Bongiorno}
\author[1]{Damien Challet}
\author[2]{Gr\'egoire Loeper}
\affil[1]{Université Paris-Saclay, CentraleSupélec,  Laboratoire de Mathématiques et Informatique pour la Complexité et les Systèmes,  91192 Gif-sur-Yvette, France}
\affil[*]{christian.bongiorno@centralesupelec.fr}

\affil[2]{BNP Paribas, 20 boulevard des Italiens, 75009 Paris, France}
\maketitle

\begin{abstract}
We propose a data-driven, model-free, way to reduce the noise of covariance matrices of time-varying systems. If the true covariance matrix is time-invariant, non-linear shrinkage of the eigenvalues is known to yield the optimal estimator for large matrices. Such a method outputs eigenvalues that are highly dependent on the inputs, as common sense suggests. When the covariance matrix is time-dependent, we show that it is generally better to use the set of eigenvalues that encode the average influence of the future on present eigenvalues resulting in a set of time-independent average eigenvalues. This situation is widespread in nature, one example being financial markets, where non-linear shrinkage remains the gold-standard filtering method. Our approach outperforms non-linear shrinkage both for the Frobenius norm distance, which is the typical loss function used on covariance filtering, and for financial portfolio variance minimization, which makes our method generically relevant to many problems of multivariate inference. Further analysis of financial data suggests that the expected overlap between past eigenvectors and future ones is systematically overestimated by methods designed for constant covariances matrices. Our method takes a simple empirical average of the eigenvector overlap matrix, which is enough to outperform non-linear shrinkage.
\end{abstract}

\section*{Introduction}
In multivariate systems, many statistical inference problems require estimating the covariance matrix or its inverse. In the simplest case, a system of interest has a constant covariance matrix and produces Gaussian features. Even in these favorable circumstances, a direct estimation of covariance matrices is very noisy when the number of data points is comparable with or smaller than the number of features, which is known as the curse of dimensionality, or high-dimensional setting. In time-evolving systems, when the underlying model is unknown, one faces the conundrum of using as few data points as possible so as to focus on the most recent information while still estimating the covariance matrix precisely enough. Thus, generically, covariance filtering requires removing two sources of systematic bias and noise: sampling noise and time evolution (which results in covariate shift).

Filtering sampling noise out of covariance matrices has a long history. There are two main approaches: either to coerce the matrix to follow a specified structure (i.e., to use an ansatz) \cite{tumminello2007hierarchically,bongiorno2020reactive},  or to use its decomposition into eigenvalues and eigenvectors (see \cite{bun2017cleaning} for a review) and filter them: for example, Random Matrix Theory provides tools to compute the noisy influence of sampling errors and reversely gives methods to denoise covariance matrices (e.g. \cite{laloux1999noise,plerou1999universal,bun2017cleaning}). Estimators that only modify the covariance matrix eigenvalues are known as Rotationally Invariant Estimators (RIE thereafter). According to recent literature, an optimal RIE should minimize the Frobenius distance between the filtered and true covariance matrix. If the true covariance matrix is known, the optimal RIE is named the Oracle estimator, and can be obtained analytically. Obviously, the Oracle estimator does not make sense for forecasting, as the true covariance is unknown. In this case,  it yields the lowest  Frobenius norm that an RIE can achieve. Remarkably, asymptotically optimal RIEs that converge to the Oracle estimator can be obtained without the knowledge of the true covariance matrix \cite{ledoit2012nonlinear,ledoit2017nonlinear,bun2016rotational,bun2017cleaning,engle2019largecov}; however, such estimators require that: i) the ground truth does not change, ii) the data matrix is very large, and iii) the data has at least finite fourth moments \cite{bun2017cleaning,ledoit2017nonlinear}. In the following, we shall call this family of estimators NLS, which stands for non-linear shrinkage.

Yet, the most interesting complex systems are rarely time-invariant and often produce heavy-tailed features. Dissipative quantum systems, ecosystems, and many socio-economic systems are time-varying in essence \cite{chen2018temporal}. Here, we propose a purely data-driven covariance filtering method that outperforms NLS as soon as the covariance matrix changes as a function of time. Remarkably, our method rests on an averaging procedure of Oracle eigenvalues in a long calibration dataset, which we call the Average Oracle. In other words, our method consists in replacing the eigenvalues of a time-dependent correlation matrix with time-invariant eigenvalues. Because the Average Oracle leads to appreciably better estimation of covariance matrices in systems with time-dependent correlation matrices, we expect its application domain to be vast. We apply it below to dynamic optimal resource allocation, which has interdisciplinary applications (financial portfolios, wind farm locations, marketing channels, and, more generally, optimization problems with a quadratic cost). 

\section*{Covariance matrix filtering}
 At time $t$, given $n$ time-series (features), one needs to predict their covariance matrix $\Sigma_\textrm{test}$ in the test interval  $\mathcal{I}_{test}=[t,t+\delta_{\textrm{test}}[$ from the information known in the train interval $\mathcal{I}_{train}=[t-\delta_{\textrm{train}},t[$. Even in the assumption of a time-invariant world, the sample estimator ${\Sigma}_\textrm{train}$  is biased and noisy as soon the ratio $q=\frac{n}{\delta_{\textrm{train}}}$ is not negligible . Fortunately, one can improve the sample estimator by using a suitable filtering scheme, which yields a new estimator, denoted by $ {\Xi}_\textrm{train}$. The idea is to bring ${\Xi}_\textrm{train}$ as close as possible to ${\Sigma}_{\textrm{test}}$, which is quantified by a distance, such as the  Frobenius distance (squared element-wise difference)  $||\Sigma_{\textrm{test}}-{\Xi}_{\textrm{train}}||_F$. 

A special class of rotationally invariant estimators uses the decomposition of the covariance matrix into eigenvectors and eigenvalues. Indeed, the spectral decomposition theorem states that $\Sigma_{train} = V_{train} \Lambda_{train} V_{train}^\dagger$, where $V_{train}$ is the $n\times n$ eigenvector matrix and $\Lambda_{train}$ is the diagonal matrix of eigenvalue monotonically ordered.

 Let us focus on eigenvalue-based filtering (i.e., build an RIE) and thus use the empirical eigenvectors $V_{\textrm{train}}$. Generically, if $\Xi$  is an RIE, it can be written as
 \begin{equation}
     \Xi(\Lambda):= V_\textrm{train} \Lambda V_\textrm{train}^\dagger, 
 \end{equation}
 where $\Lambda$ is a diagonal matrix with well-chosen eigenvalues. For example, if one knows the future covariance matrix  $\Sigma_\textrm{test}$, the optimal eigenvalue matrix is the so-called Oracle and can be shown \cite{bun2016rotational} to be   
\begin{equation}\label{eq:oracle}
\Lambda_{O}=\textrm{diag}(  V_{\textrm{train}}^{\dagger} \Sigma_{\textrm{test}} V_{\textrm{train}}),
\end{equation}
where the $\textrm{diag}$ operator only keeps the diagonal of a matrix and sets the elements to zero elsewhere. These eigenvalues express the future empirical covariance matrix in the basis of the current one. They are optimal in the following sense: the related RIE
\begin{equation}
{\Xi}(\Lambda_{O})= V_{\textrm{train}} \Lambda_{O} V_{\textrm{train}}^{\dagger},
\end{equation}
minimizes the (element-wise) Frobenius distance  $||{\Xi}(\Lambda_{O})-{\Sigma}_{\textrm{test}}||_F$. 
 Although the exact Oracle estimator cannot be used for practical purposes, since $\Sigma_\textrm{test}$ is in the future, asymptotical estimators that converge to the Oracle estimator are known \cite{ledoit2012nonlinear,bartz2016cross,ledoit2017nonlinear,bun2017cleaning}. In other words, this optimal RIE exploits a way to express $\Lambda_O$ as a function of the past information only, provided that the hypotheses listed above hold, temporal independence being a crucial one.

The above asymptotically optimal RIEs rest on neglecting any evolution of the covariance matrix. Although this makes sense for constant underlying covariance matrices, it likely discards relevant information for time-dependent matrices, and indeed  Oracle eigenvalues do contain valuable information to filter ${\Sigma}_{\textrm{train}}$ as they encode the link between the past and the future in an RIE setting. 
With the aim of capturing the average transition from two consecutive time windows, our method rests on the averaging Eq.\ \eqref{eq:oracle}, rank-wise, over many randomly selected consecutive intervals taken from a long calibration window. The latter must be much larger than the one used to compute $V_{\textrm{train}}$.

\begin{figure}
    \centering
    \begin{tikzpicture}
\draw[->, thick] (-5,0)--(5,0) node[right]{time};
\draw[decoration={brace,raise=5pt},decorate]
  (-4.5,0) -- node[above=8pt] {$\mathcal{I}_\textrm{prev}^{(2)}$} (-3.25,0);
\draw[black, thick] (-3.25,-0.1) -- node[below=8pt]{$t^{(2)}$} (-3.25,0.1);
\draw[decoration={brace,raise=5pt},decorate]
  (-3.25,0) -- node[above=8pt] {$\mathcal{I}_\textrm{next}^{(2)}$} (-2.25,0);
  
\draw[decoration={brace,raise=5pt},decorate]
  (0,0) -- node[above=8pt]
  {$\mathcal{I}_\textrm{prev}^{(1)}$} (1.25,0);
\draw[black, thick] (1.25,-0.1) -- node[below=8pt]{$t^{(1)}$} (1.25,0.1);
\draw[decoration={brace,raise=5pt},decorate]
  (1.25,0) -- node[above=8pt] {$\mathcal{I}_\textrm{next}^{(1)}$} (2.25,0);

\draw[decoration={brace,mirror,raise=5pt},decorate]
  (-2.9,0) -- node[below=8pt]
  {$\mathcal{I}_\textrm{prev}^{(3)}$} (-1.65,0);
\draw[black, thick] (-1.65,-0.1) -- node[above=8pt]{$t^{(3)}$} (-1.65,0.1);
\draw[decoration={brace,mirror,raise=5pt},decorate]
  (-1.65,0) -- node[below=8pt] {$\mathcal{I}_\textrm{next}^{(3)}$} (-0.65,0);

\draw[black, thick] (3,-0.1) -- node[above=8pt] {$t$}(3,0.1);
\draw[decoration={brace,mirror,raise=5pt},decorate]
  (1.75,0) -- node[below=8pt] {$\mathcal{I}_\textrm{train}$} (3.,0);
\draw[decoration={brace,mirror,raise=5pt},decorate]
  (3.,0) -- node[below=8pt] {$\mathcal{I}_\textrm{test}$} (4,0);

\draw[decoration={brace,raise=5pt},decorate]
  (-5,3) -- node[above=8pt] {Calibration window $\mathcal{I}_\textrm{cal}$} (2,3);

\draw[black, dotted] (3,0) -- node[above=45pt] {today}(3,3);
\end{tikzpicture}
    \caption{Average Oracle eigenvalues computation: Oracle eigenvalues are computed in many sub-intervals of a long calibration window and then averaged rank-wise. These eigenvalues can then be used outside of the calibration window.}
    \label{fig:AO_tikz}
\end{figure}
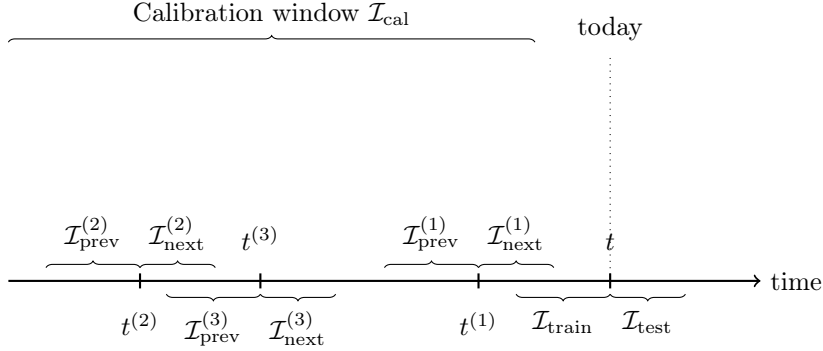

More precisely (see Fig. \ref{fig:AO_tikz}  for a graphical explanation), we need to define a long calibration window $\mathcal{I}_{cal}=[t-\delta_{\max}, t-\delta[$, with $\delta_{\max} \gg \delta_{train}$,  and $\delta$  not necessarily linked to the actual test window size $\delta_\textrm{test}$, as shown in appendix \ref{sec:delta_test}. Then, we select $B$ random times $t^{(b)}\in \mathcal{I}_{cal}$. For a given $t^{(b)}$, two consecutive intervals must defined: the first interval $\mathcal{I}^{(b)}_{prev}=[t^{(b)}-\delta_\textrm{train},t^{(b)}[$ must be of size $\delta_\textrm{train}$, while the length of the next one  $\mathcal{I}^{(b)}_{next}=[t^{(b)}, t^{(b)}+\delta[$ will be of size $\delta$. It is worth mentioning that the next interval $\mathcal{I}^{(b)}_{next}$ is in the future with respect $t^{(b)}$ but it is in the past with respect $t$, i.e., with respect $\mathcal{I}_{test}$; therefore, an Oracle-like scheme can be applied. 

Each sub-sampling has an associated set of Oracle eigenvalues
\begin{equation}\label{eq:oracboot}
\Lambda^{(b)}_{O}=\textrm{diag}(  {V_{\textrm{prev}}^{(b)}}^{\dagger} \Sigma^{(b)}_{\textrm{next}} V_{\textrm{prev}}^{(b)}).
\end{equation}
With $V_{\textrm{prev}}$ the eigenvectors of the sample covariance matrix computed in $\mathcal{I}^{(b)}_{prev}$, and $\Sigma^{(b)}_{\textrm{next}}$ the sample covariance of $\mathcal{I}^{(b)}_{next}$. 
The Average Oracle eigenvalues are then defined as the average element-wise:
\begin{equation}
\Lambda_{AO}:=\frac{1}{B}\sum_{b=1}^B\Lambda^{(b)}_{O}.
\end{equation}
It is important to stress that the columns of the eigenvectors $V_{\textrm{prev}}^{(b)}$  of \eqref{eq:oracboot} must always follow the same the eigenvalue ordering convention chosen.

The AO-filtered covariance matrix is given by 
\begin{equation}
    \label{eq:sigma_AO}
\Xi(\Lambda_{AO})=V_{\textrm{train}}\Lambda_{AO}V_{\textrm{train}}^\dagger.
\end{equation}
The empirical eigenvalues from the train interval are completely discarded and replaced by the AO ones. The fact that the Average Oracle is a better estimator for time-evolving covariance matrices most often implies that the most recent information contained in the sample eigenvalues is less relevant (and more noisy) than the AO ones that focus on the average transition. On the other hand, the train eigenvectors contain some dynamical information and are kept. Note also that our approach requires that the univariate variances are constant. If one deals with a system in which this is not the case, our method should be applied to the correlation matrix.

 We thus propose to tackle the evolution of dependencies with a time-invariant eigenvalue cleaning scheme. This is a zeroth order approximation, as the fluctuations of the optimal eigenvalue matrix around $\Lambda_{AO}$  sometimes most probably contain valuable additional information (as may do those of the eigenvectors). Nevertheless, this approximation is a powerful filtering tool and is easily computed from data without any modeling assumptions about the underlying system. In addition, our tests indicate that the filtering power does not decrease substantially as a function of time in the systems that we investigated. 


\section*{The Role of the Temporal Evolution}
To understand how the Oracle eigenvalues are related to the time-dependency of both the eigenvalues and eigenvectors of a covariance matrix, we decompose \eqref{eq:oracle} as
\begin{equation}\label{eq:oracleoverlapmat}
    \lambda_{O} = \left( V_{\textrm{prev}}^\dagger V_{\textrm{next}}\right)^{\circ 2} \lambda_{\textrm{next}} = H^{\circ 2} \lambda_{\textrm{next}},
\end{equation}
where $\lambda_{O}$ and $\lambda_{\textrm{next}}$ are the eigenvalues shaped in column vectors, and $\bullet^{\circ 2}$ represents the Hadamard element-wise square of the matrix $H$.  The matrix $H$  is in fact a rotation matrix from the $V_\textrm{prev}$ eigenvector basis to the $V_\textrm{next}$ eigenvector basis and hence belongs in $SO(n)$.

\begin{figure}
    \centering
    \includegraphics[width=8.0cm]{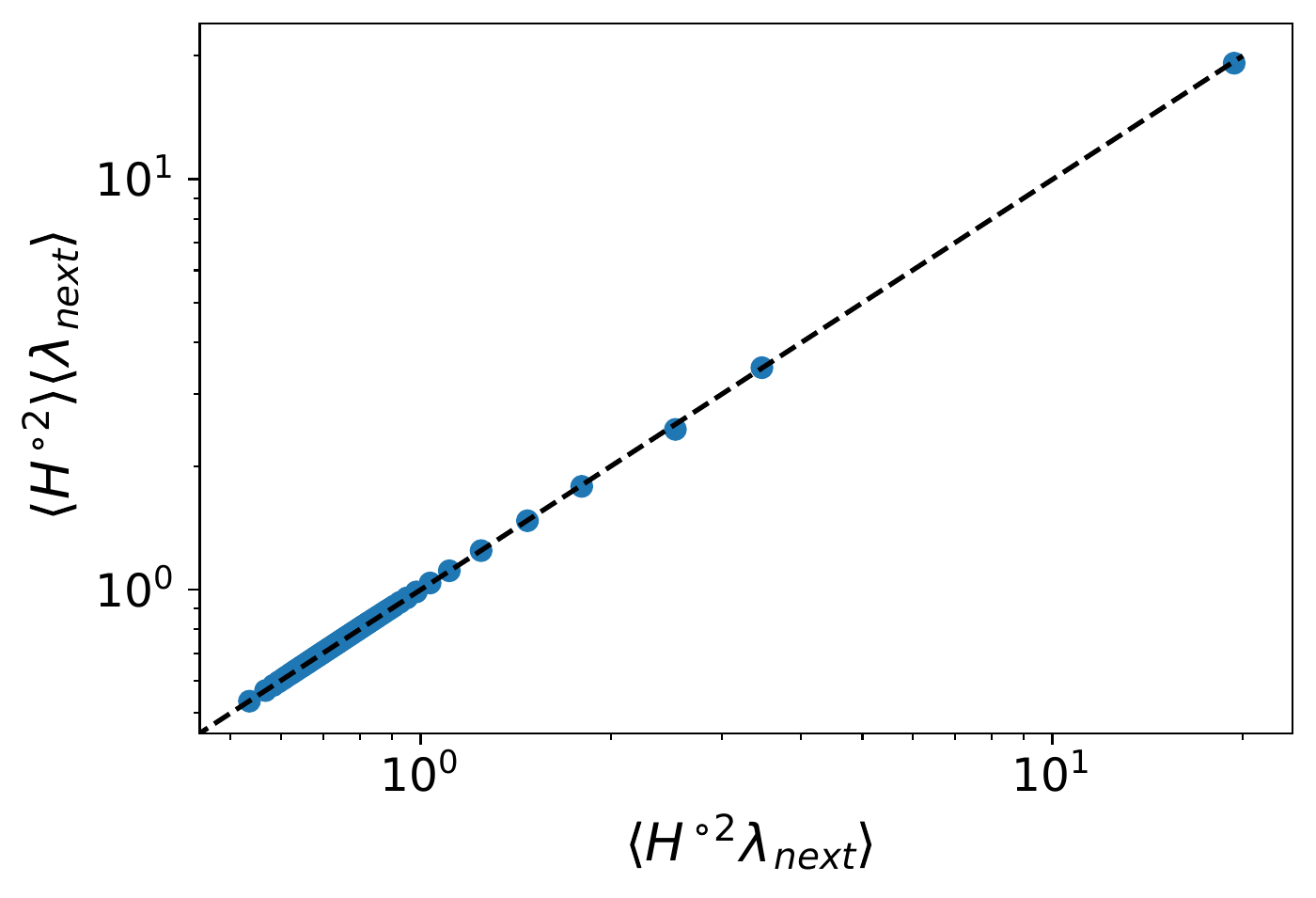}
 \caption{Empirical test of the linear independence of $\lambda_{\textrm{next}}$ and $H^{\circ 2}$ (Eq.~\eqref{eq:orsepavg}): $\langle H^{\circ 2} \rangle \langle \lambda_{\textrm{next}} \rangle$ vs $\langle H^{\circ 2} \lambda_{\textrm{next}} \rangle$.}
    \label{fig:statInd}
\end{figure}  

This decomposition of the Oracle estimator makes it possible to analyze the respective contributions of $H$ and $\lambda_{\textrm{next}}$ to the Oracle estimator and, by linearity, to the Average Oracle. Indeed, according to Eq.~\ref{eq:oracleoverlapmat}, it is natural to test if the problem of overlap and test eigenvalues estimation can be separated, i.e., if
\begin{equation}\label{eq:orsepavg}
    \langle H^{\circ 2} \lambda_{\textrm{next}} \rangle \simeq \langle H^{\circ 2} \rangle \langle \lambda_{\textrm{next}} \rangle.
\end{equation}
It is worth noticing that although the average element-wise of an element of the $SO(n)$ group is not an element of the $SO(n)$ group, the element-wise average of $H^{\circ 2}$ still has the same relevant property as the original elements, i.e., the row-wise or column-wise sum equal one. 

We test Eq.~\eqref{eq:orsepavg} with financial data. It is worth stressing, as detailed in the appendix \ Ref {sec:variance}, that the variance of each time series varied tremendously in the last twenty years and hence we focus on correlation rather than covariance matrix. In practice, we standardize the time series on every subinterval considered  (see the appendix \ref{sec:data} for a full description of data handling).

We find that Eq.~\eqref{eq:orsepavg} holds remarkably well (Fig.~\ref{fig:statInd}):  $H^{\circ 2}$ and  $\lambda_{\textrm{next}}$ are linearly independent and thus both quantities can be assumed to contribute independently to the fluctuations of the Oracle eigenvalues. As a consequence, we are allowed to focus on the influence of eigenvalues and eigenvectors separately.

\subsection*{Overlap matrix stability}
  The element-wise square $h_{ij}^2$ represents the projection of the $i$ eigenvector from $V_{\textrm{prev}}$ on the eigenvector $j$ from $V_{\textrm{next}}$ . In the case of the perfect overlap, only one $h_{ij}=1$, all the others being zero. This never happens because of finite-sample size error and temporal evolution. The other extreme case is  $h_{ij}^2=\frac{1}{n}$ for all $j$, which corresponds to the lowest overlap possible. Given these mathematical properties of the overlap matrix,  Shannon entropy is an appropriate measure of the amount of overlap. For eigenvector $i$, we write
\begin{equation}\label{eq:vectoverlap}
    E_i = - \sum_{j=1}^n h_{ij}^2 \log_n h_{ij}^2
\end{equation}
with the standard  Shannon entropy convention that $0 \log 0 = 0$ and normalization by $\log(n)$, in such a way that the highest overlap will be $E_i=0$ while the lowest $E_i=1$. 

By using Eq.~\eqref{eq:vectoverlap}, we can test on real data if the average overlap in a time-invariant world, due only to sampling size error, is compatible with the measured overlap of the real world. We carried out such an experiment in the following way. We randomly select $10,000$ random time-windows  $\mathcal{I}^{(b)}_{\textrm{prev}}$, $\mathcal{I}^{(b)}_{\textrm{next}}$ with $\delta_{prev}=\delta=252$ days and perform an independent random selection of $n$ stocks for each time-window.  We then measure the entropy of the overlap for each $V_{\textrm{prev}}$ eigenvector ranked from the smallest eigenvalue to the largest one. The entropy is then averaged element-wise over these independent realizations, which yields a measure of real-world overlap, influenced both by sampling error and time evolution. 

\label{sec:stat}
To show that the real-life overlap is systematically lower than the one expected for a time-invariant world, we design a local data shuffling procedure to remove any effect of temporal evolution, and we measure the rank-wise average overlap $E_i$ again. First we take the union  of the intervals $\mathcal{I}^{(b)}_{\textrm{prev}} \cup \mathcal{I}^{(b)}_{\textrm{next}}$, then we shuffle the temporal ordering of the observations,  then we split the shuffled data again into  $\mathcal{I}'^{(b)}_{\textrm{prev}}$, $\mathcal{I}'^{(b)}_{\textrm{next}}$, but after the shuffling both intervals will contain a mixture a past and future events; therefore, any quantity has the same expectation in both intervals.

\begin{figure}
    \centering
    \subfigure[]{\includegraphics[width=8.0cm]{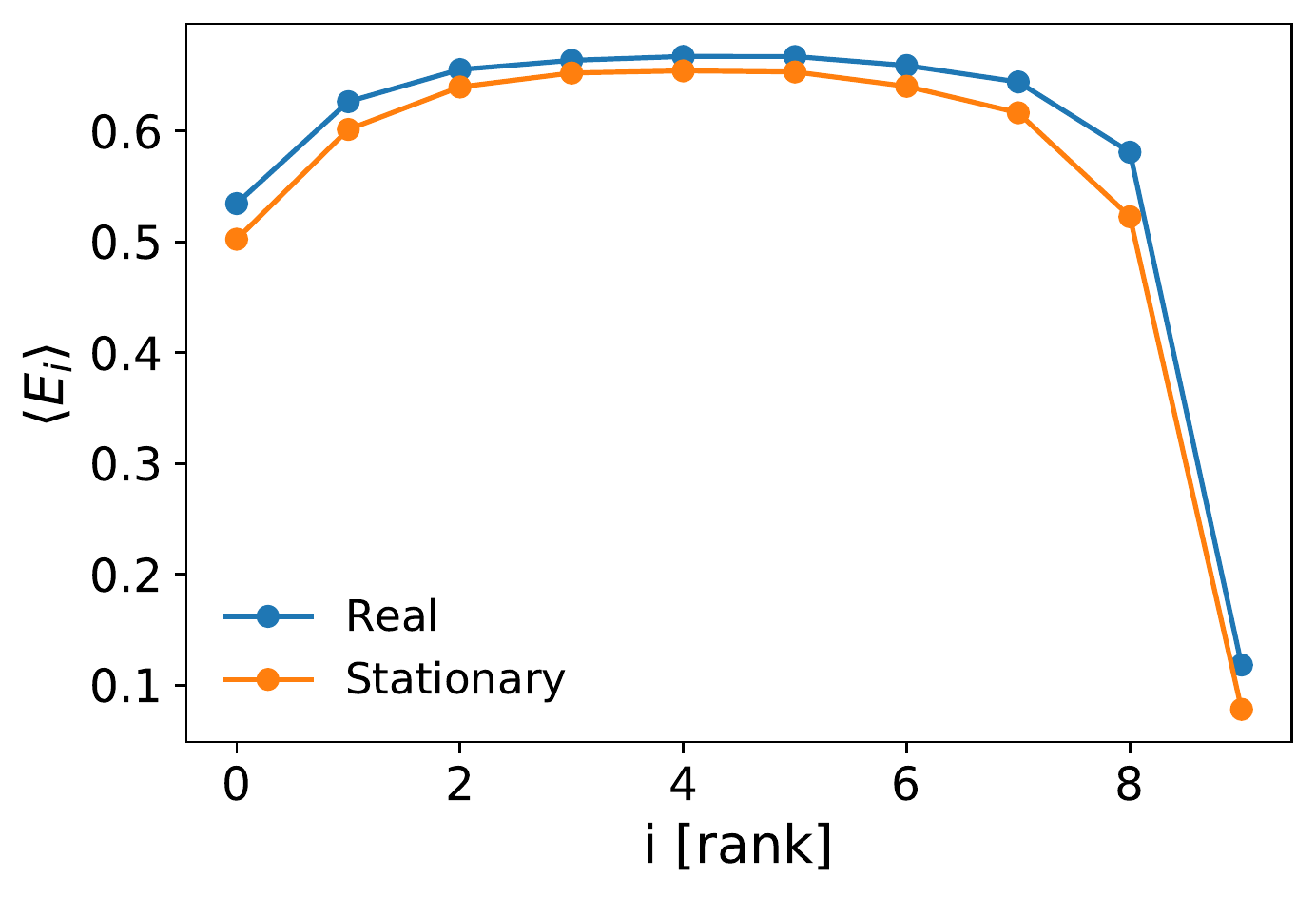}}
\subfigure[]{   \includegraphics[width=8.0cm]{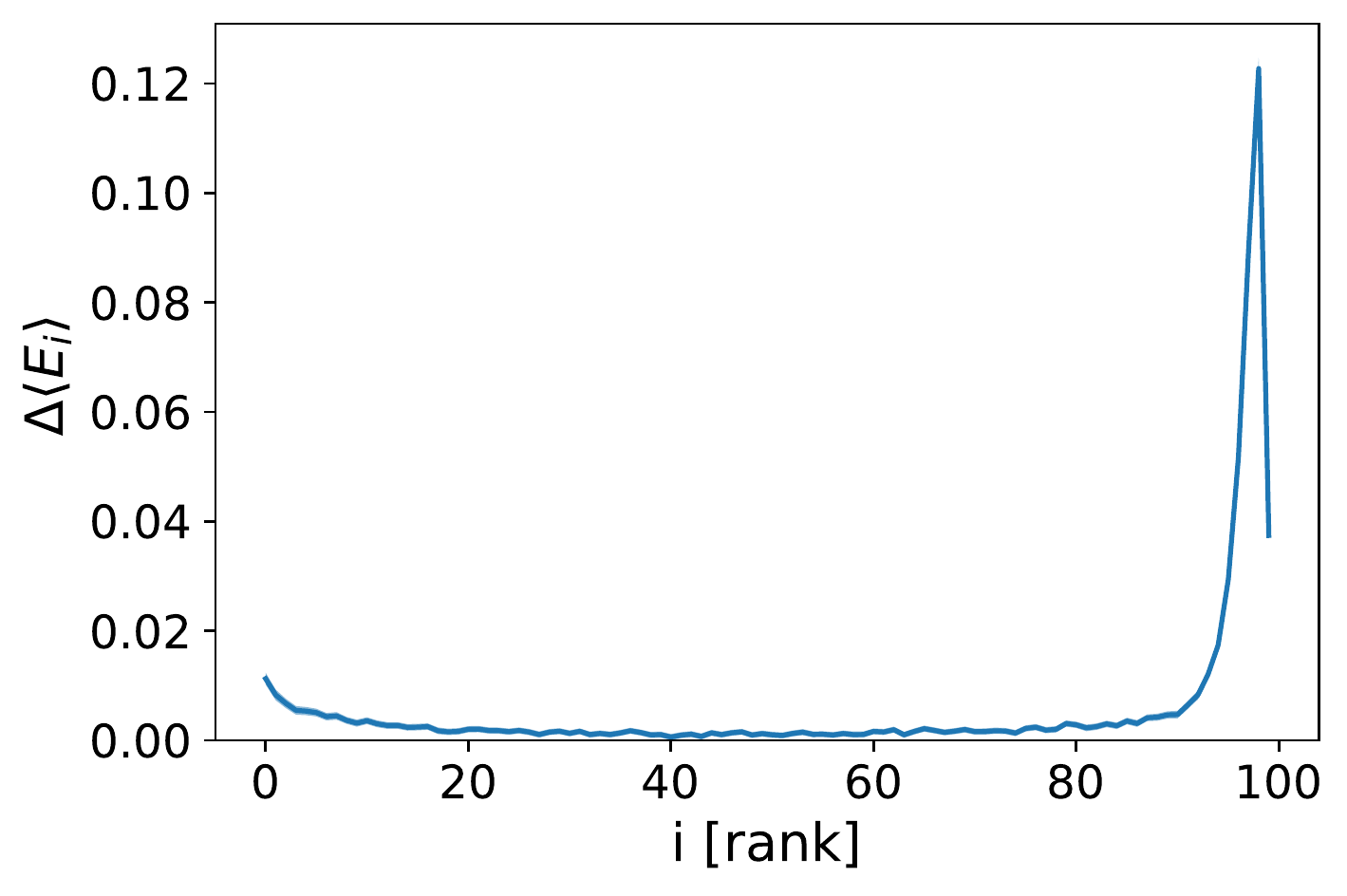}}
     \caption{Entropy of the eigenvector overlap for  US equities. Each point is an average of $10,000$ independent realizations. The upper plot refers to $n=10$; the lower plot is the difference between the real-life entropy and the shuffled data entropy for $n=100$; the plot includes $95\%$ confidence intervals which are not clearly visible due to the precision of the estimation. }
    \label{fig:entropy}
\end{figure}  
 
 Figure~\ref{fig:entropy} displays the average entropy for both cases. One sees that data shuffling leads to smaller entropy, hence, larger apparent overlap. This means that in a fictional time-invariant world, the overlap is mechanically larger, which in turn implies that this assumption leads to a bias in the eigenvectors' stability in the real world. While for $n=10$ the overestimation of the overlap for the time-invariant cases is clear, for $n=100$ we must look at the difference between the two estimators. One notes that the overestimation of the time-invariant case is systematic even for large $n$.
 
\subsection*{Eigenvalues stability}
In Eq.~\ref{eq:oracle}, the other part of Oracle the estimator is the expectation of $E[\Lambda_\textrm{next}]$. A reasonable assumption in a time-invariant world, if $\delta_\textrm{train}=\delta$, is that $E[\Lambda_\textrm{next}] = \Lambda_\textrm{prev}$. 

Another possibility is that, because of very fast temporal evolution, the eigenvalues fluctuate so much that their  average  $\langle \Lambda_\textrm{next} \rangle$ computed over many time-windows within a much larger calibration time-window $\mathcal{I}_{\textrm{cal}}$ is a better predictor of $\Lambda_\textrm{next}$ than the closest past $\mathcal{I}_{\textrm{prev}}$.

In order to test these two hypotheses, we computed the average $\langle \Lambda_\textrm{next} \rangle$ over $10,000$ randomly chosen time-windows of $\delta_{prev}=\delta=252$ days in the the calibration window $[1995,2005]$. We then we tested the deviation $\Lambda_\textrm{prev}-\Lambda_\textrm{next}$ and $\langle \Lambda_\textrm{next}\rangle - \Lambda_\textrm{next}$ in the test window $[2006,2018]$ with $10,000$ randomly chosen time-windows. We carried out such a comparison with $L_1$ and $L_2$ norms.

Specifically, the $L_1$ norm is defined as
\begin{equation}
    D^{(1)}_\bullet = \sum_{i=1}^n | \lambda_{i,\bullet} - \lambda_{i,\textrm{next}}|
\end{equation}
and the $L_2$ norm is defined as
\begin{equation}
     D^{(2)}_\bullet =\sqrt{\sum_{i=1}^n\left(\lambda_{i,\bullet} - \lambda_{i,\textrm{next}}\right)^2}
\end{equation}
where $\lambda_{i,\bullet}$ can be the $\langle \lambda_{i,\textrm{next}}\rangle$  or $\lambda_{i,\textrm{prev}}$. 

\begin{figure}
    \centering
 \subfigure[]{   \includegraphics[width=8.0cm]{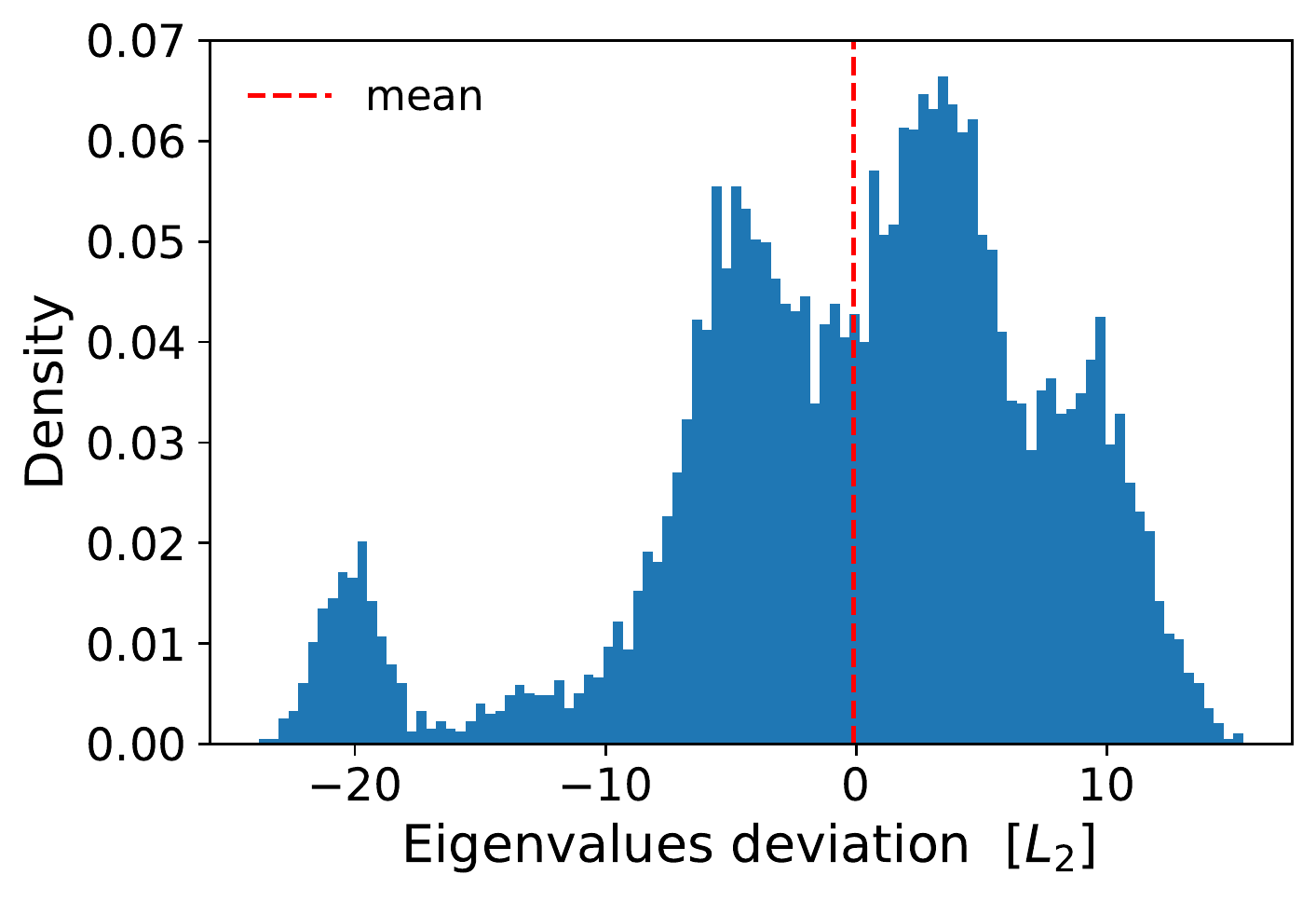}}
\subfigure[]{   \includegraphics[width=8.0cm]{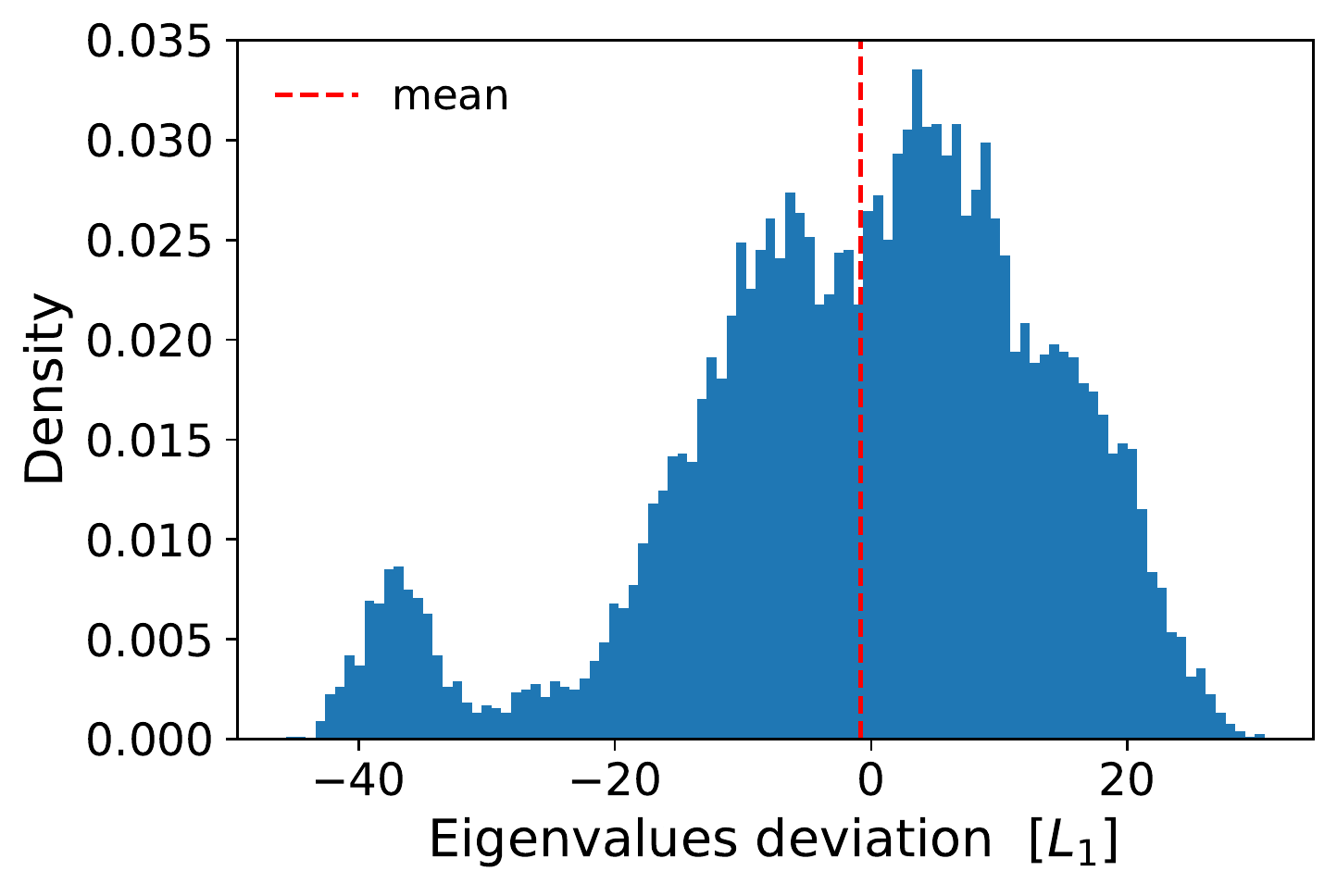}}

 \caption{Plot $(a)$: distribution of eigenvalue deviation $D^{(2)}_{\langle \textrm{next} \rangle}$-$D^{(2)}_{\textrm{prev}}$; plot $(b)$: distribution of eigenvalue deviation $D^{(1)}_{\langle \textrm{next} \rangle}$-$D^{(1)}_{\textrm{prev}}$; each element of the distribution refers to a randomly sampled time-window and a random set of $n=100$ stocks; $\delta_{prev}=\delta=252$.}
    \label{fig:eigdev}
\end{figure}

In Fig.~\ref{fig:eigdev}, we show the distribution of the eigenvalue deviation as $D^{(2)}_{\langle \textrm{next} \rangle}$-$D^{(2)}_{\textrm{prev}}$ and $D^{(1)}_{\langle \textrm{next} \rangle}$-$D^{(1)}_{\textrm{prev}}$. Each point of the distribution is a random selection of the time window on the validation period.
The average of both distributions is close to zero, supporting the idea that the most recent eigenvalues are approximately as good as the average historical ones. More precisely the distribution average is slightly negative: $-0.1$ for the $L_2$, with a 95-percentile bootstrap p-value of $0.1$ and $-0.8$ with a 95-percentile bootstrap p-value of zero (computed with $100,000$ bootstrap re-sampling). However, such a difference is marginal, and the average approach does not bring any significant improvement.

\section*{Matrix distance between filtered and realized covariance}

We expect our method to be useful for any system with time-dependent dependencies as a zeroth order correction. Here, we test AO vs.~NLS with financial data as they are abundant, have dynamic dependencies, and are heavy-tailed \cite{gopikrishnan1999scaling}. We also use synthetic data to assess the respective influence of covariance temporal evolution and heavy tails on the performance of both AO and NLS in appendix \ref{sec:model_nonstat}.

In some systems with time-dependent covariances, such as financial markets, the order of magnitude of univariate variances $\Sigma_{ii}$ strongly depends on time. In order to remove this source of time-dependence, we focus on the eigenvalue correction of the correlation matrix and compute $\Xi(\Lambda_{AO})=V_{\textrm{train}}^\dagger\Lambda_{AO}V_{\textrm{train}}$ where the eigenvalues and eigenvectors are those of correlation matrices. The elements of a filtered covariance matrix are obtained by multiplying the filtered correlation matrix by the respective univariate standard deviations
\begin{equation}
(\Sigma_{AO})_{ij}=\sqrt{(\Sigma_\textrm{train})_{ii}}\Xi(\Lambda_{AO})_{ij}\sqrt{(\Sigma_\textrm{train})_{jj}}\label{eq:cov_corr}
\end{equation}

\begin{figure}
    \centering
    \includegraphics[width=8.6cm]{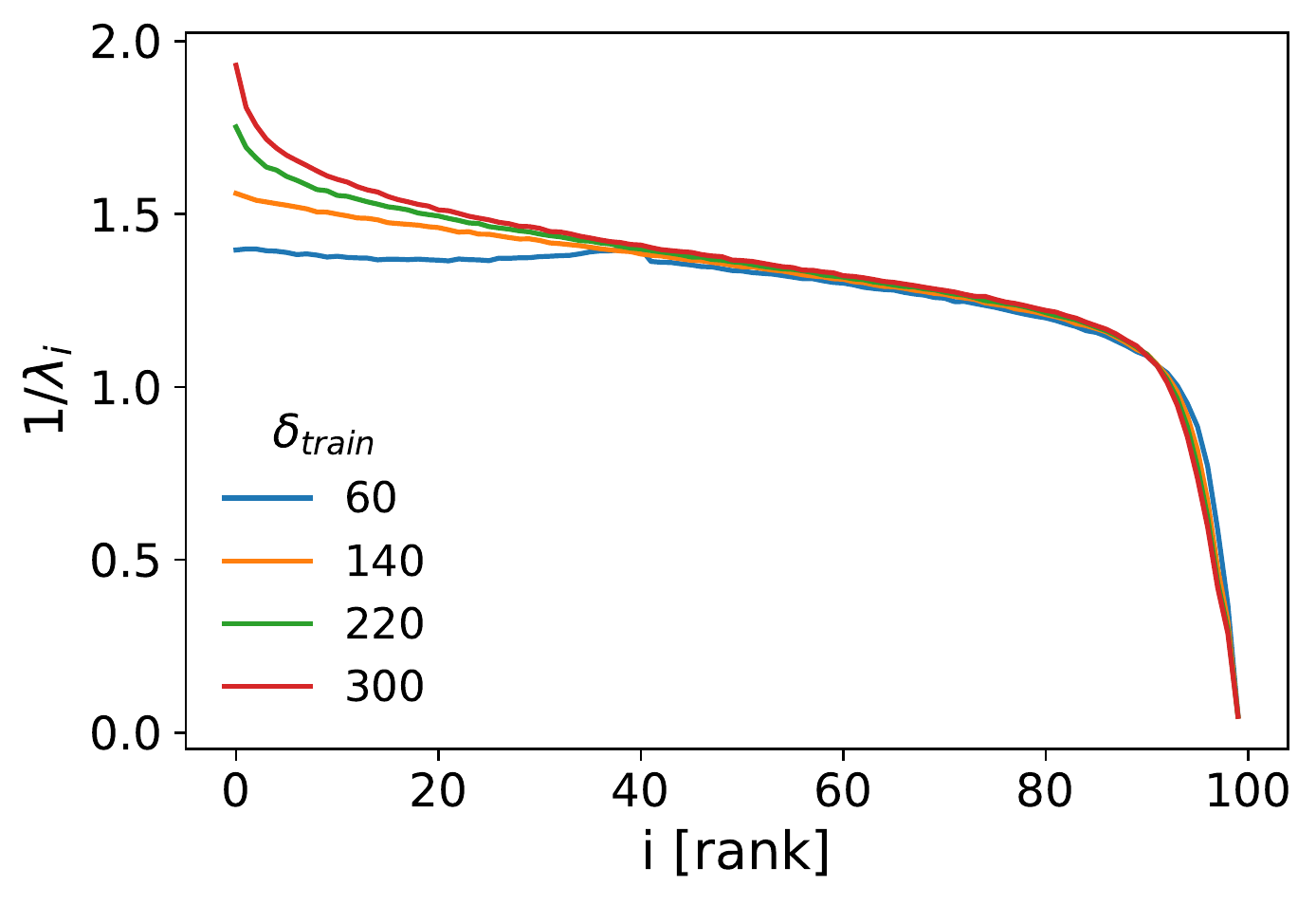}
    \caption{Inverse Average Oracle eigenvalues as a function of the eigenvalue rank for $n=100$ and various train window sizes. US Equities correlation matrix; $B=10000$ sub-intervals in the 1995-2006 period.}
    \label{fig:avg_lambda_k}
\end{figure}

In the following, we use about 24 years of daily data for $N=1000$ assets from the US stock market. The Average Oracle eigenvalues $\Lambda_{AO}$ are calibrated in the 1995 to 2005 period from $B=10000$ subintervals; for the sake of computation speed, we take $n<N$ random assets in each subinterval. Because we randomize asset selection, the resulting Average Oracle eigenvalues can be applied to any selection of assets (and to other markets). One could also choose to compute $\Lambda_{AO}$ for a fixed set of assets. For a full description of the data handling, see the appendix \ref{sec:data}. 

The resulting AO eigenvalues are reported in Fig.\ \ref{fig:avg_lambda_k}. Note that we plot the inverse of the average eigenvalues in order to emphasize the dependence on $n$ of small eigenvalues, as many inference problems use the inverse covariance matrix (see below), and also to reduce the influence of the largest eigenvalues on the clarity of the figure.

\begin{figure}
    \centering
    \subfigure[]{\includegraphics[width=8.cm]{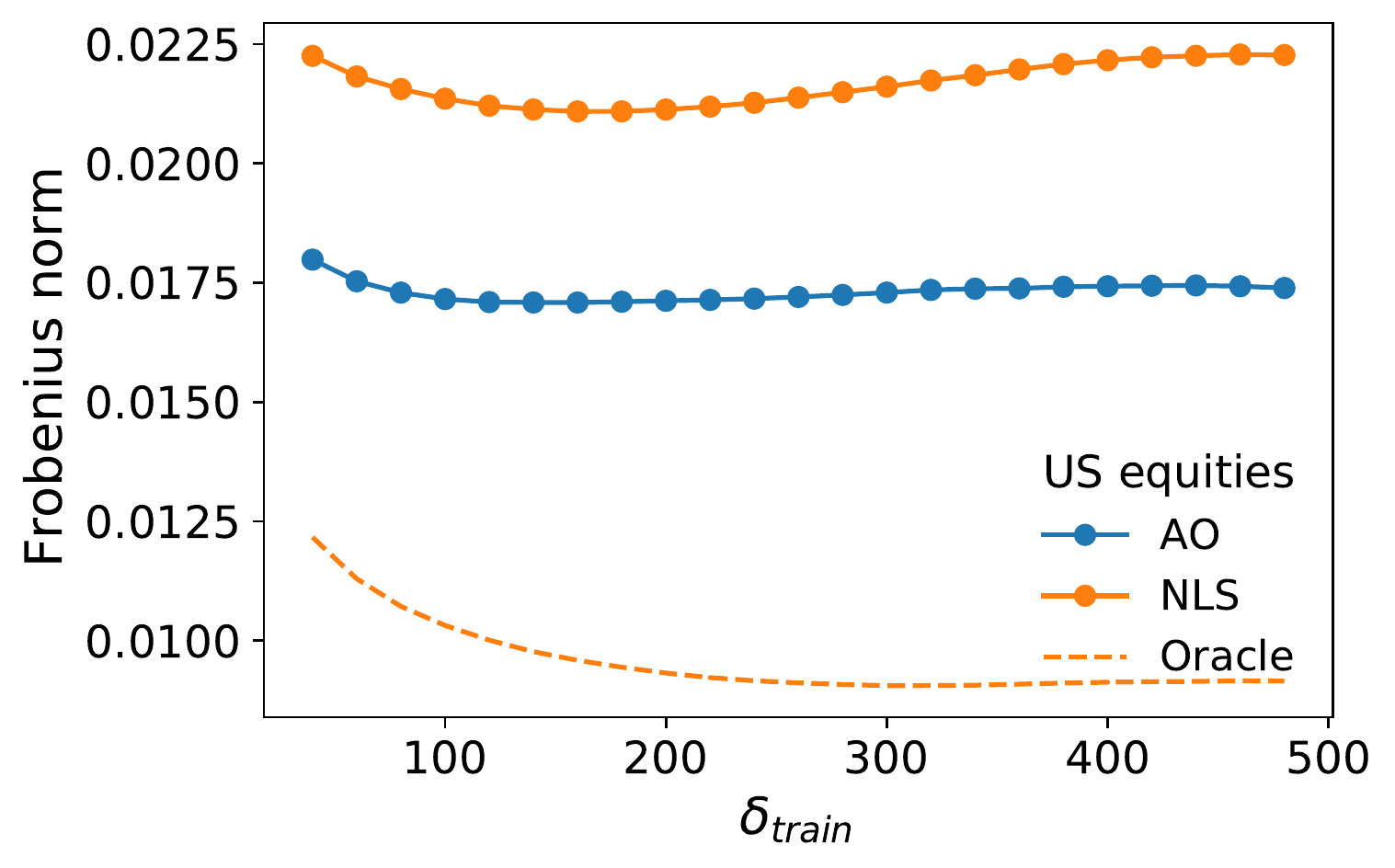}}
     \subfigure[]{\includegraphics[width=8.cm]{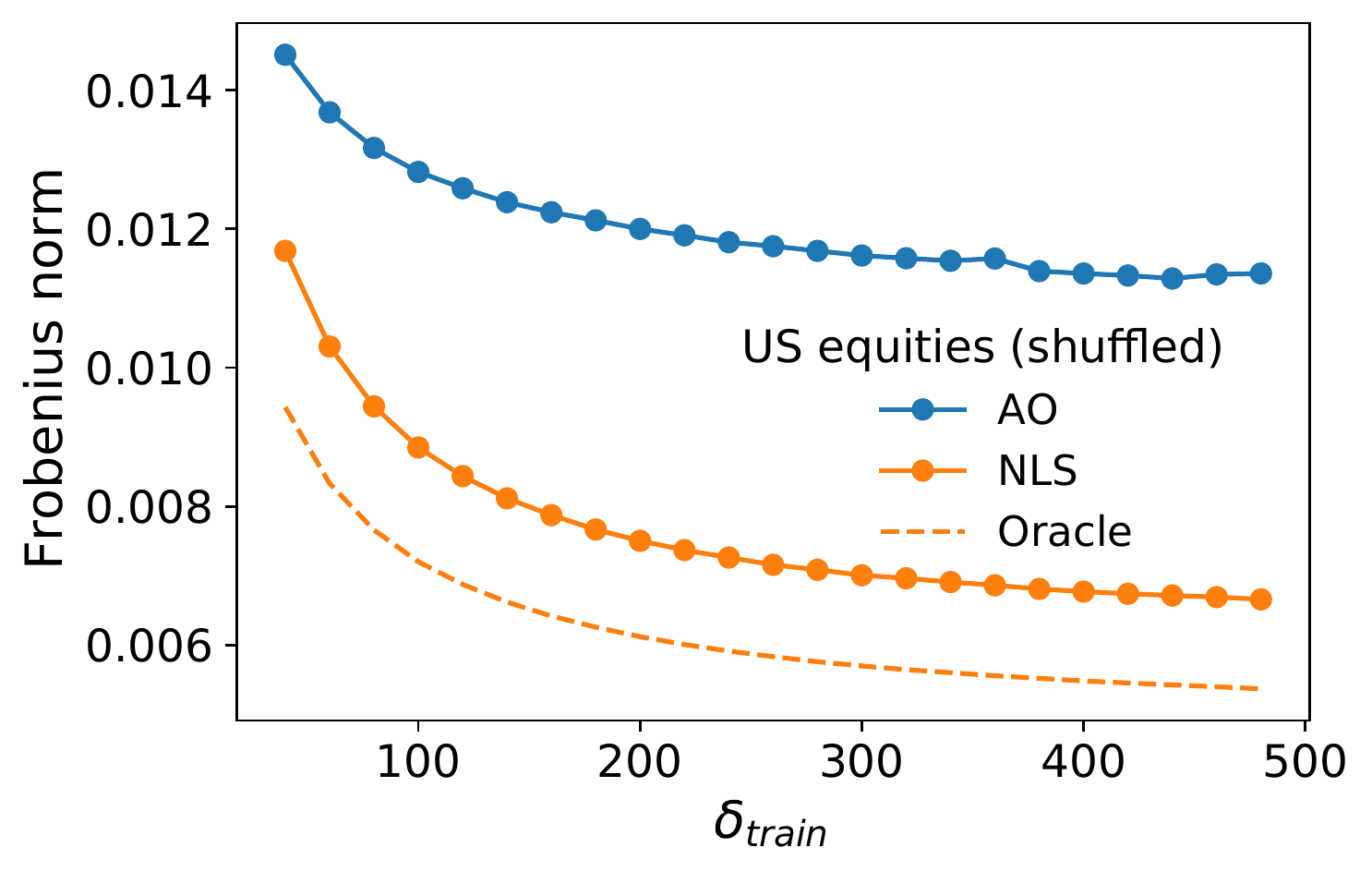}}
    \caption{Average Frobenius distance between the filtered and test covariance matrices as a function of the train window length in the out-of-sample period. Plot $(a)$ refers to the original data set, and plot  $(b)$  to the shuffled data. 100 portfolios with random $n=100$ assets are computed for each day of the out-of-sample.} 
    \label{fig:Frobenius_RU}
\end{figure}

The first way to compare the performance of both NLS and AO is to compute the average Frobenius norm in the out-of-sample period ($2006-2018$). 
We carried out extensive simulations with various $\delta_{\textrm{train}},\delta_{\textrm{test}}\in\{40,\cdots,500\}$, selecting 100 random selections of $n=100$ assets for each  $t^{(b)}$. While $\Lambda^{AO}$ depends on $n$ and $\delta_\textrm{train}$, it does not seem to depend on $\delta$ (see appendix \ref{sec:delta_test}) which we fix henceforth to 252 (one year of daily data).  We compared the Average Oracle approach with an efficient and provably good numerical implementation of NLS \cite{bartz2016cross,bun2018overlaps} based on cross-validation within the train window. 

The average Frobenius norm in the out-of-sample period for NLS and AO is reported in the left plot of Fig.\ \ref{fig:Frobenius_RU}. AO clearly does better than NLS, even if the latter is designed to minimize this norm in the time-invariant case. For the sake of completeness, we also added the unrealistic case where the Oracle eigenvalues are computed from the future as in Eq.\ \eqref{eq:oracle}, which shows how much the AO could still be improved with a predictive model of eigenvalues. Note that the advantage of AO over NLS increases with $n$ (appendix \ref{sec:test_windowlength}). We also check in appendix \ref{sec:KL} that the same results hold for the Kullback-Leibler distance.

The superior performance of the Average Oracle is mainly due to the time evolution of the true correlation matrix. Indeed, when we shuffle the data in the train and test interval as described in Sec.~\ref{sec:stat}, the advantage of using AO disappears. In particular,  in Fig.\ \ref{fig:Frobenius_RU} (right plot), we show that NLS clearly outperforms the Average Oracle on the Frobenius distance on shuffled data. Thus, the advantage of the Average Oracle is precisely that it captures some part of the average dynamics that is discarded by the assumption of a constant true covariance matrix (see also appendix \ref{sec:model_nonstat}).


\section*{Application to portfolio optimization}

\begin{figure}
    \centering
    \subfigure[]{\includegraphics[width=8.cm]{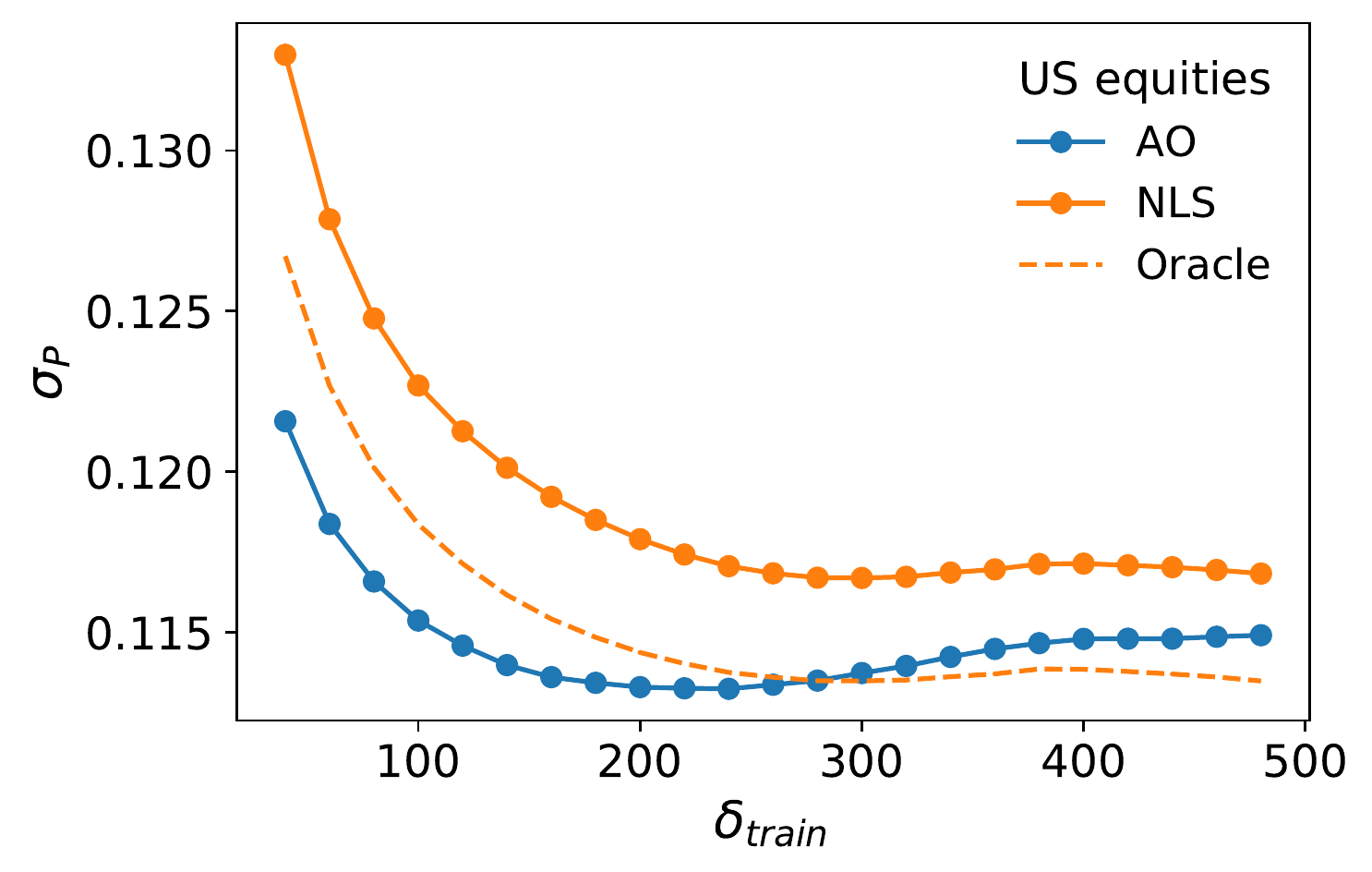}}
    \subfigure[]{\includegraphics[width=8.cm]{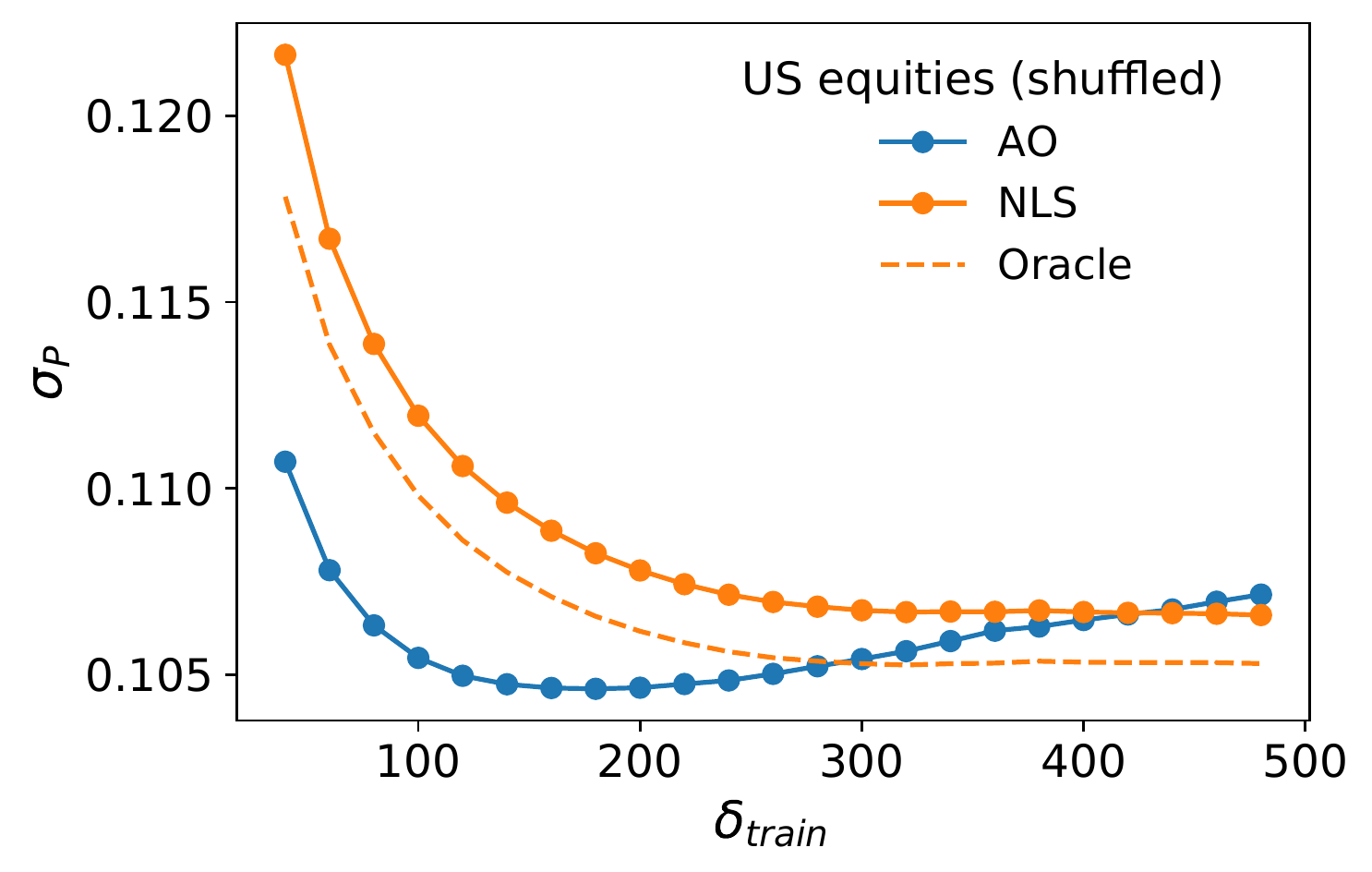}}
    \caption{Average realized volatility of Global Minimum Variance portfolios as a function of the calibration window length. Plot $(a)$ refers to the original dataset, and plot $(b)$ to the shuffled data. 100 portfolios with random $n=100$ assets are computed for each day of the out-of-sample.}
    \label{fig:vol_RU}
\end{figure}

Portfolio optimization is a canonical application of covariance matrices in a resource allocation context. The simplest case only uses the covariance matrix and aims at minimizing the realized variance of the value of a portfolio of assets from the knowledge of data in the train interval. Mathematically, a portfolio is defined by the fraction $w_i$ of wealth invested into each available asset $i=1,\cdots,n$. In other words, the performance of a portfolio with weights $w\in\mathbb{R}^n$ is the weighted sum of the performance of all the assets, i.e., $r_P=\sum_{i=1}^n w_i r_i$, where $r_i$ is the price return of asset $i$, and its variance is ${w}^\dagger\Sigma w$.  Practically, the weights are computed from the train window data, and the covariance is that of the test window, thus the realized portfolio volatility $\sigma_P$ is given by
\begin{equation}
 \left(\sigma_P\right)^2={w}^\dagger {{\Sigma}}_{\textrm{test}}{w}.
\end{equation}

The optimization problem usually adds the normalization constraint $\sum_{i=1}^N w_i=1$. This defines the Global Minimum Variance Portfolio problem (GMV). 
Intuitively, minimizing $\sigma_P$ requires a small distance between $\Sigma_{train}$ and $\Sigma_{test}$, hence the importance of the Frobenius norm (see Fig.\ \ref{fig:Frobenius_RU}), which is the usual criterion in the covariance matrix filtering literature \cite{bun2017cleaning}, which however, does not lead to the optimal portfolio weights \cite{bongiorno2021oracle} in finite sample size cases. 

The weights that minimize the portfolio variance given a covariance estimator $\Sigma$ can be computed analytically 
\begin{equation}
    {w}^*=\frac{{\Sigma}^{-1}\mathbb{1}}{\mathbb{1}^\dagger{\Sigma}^{-1}\mathbb{1}}.
\end{equation}
One immediately notices that portfolio optimization also requires that the inverse of the covariance matrix (the precision matrix) be also well filtered as the optimal weights are much influenced by the smallest eigenvalues of $\Sigma$. 

Figure \ Ref.~\ref{fig:vol_RU} shows that the realized volatility of GMV portfolios is smaller when using the Average Oracle than when using NLS, as expected from the Frobenius norm. However, quite remarkably, AO is also better than the Oracle eigenvalues when $\delta_{\textrm{train}}<270$. 
We use the same shuffling procedure of the previous section to check the importance of correlation time evolution. The Average Oracle still outperforms the Oracle and NLS for small enough $\delta_{\textrm{train}}$. i.e., in the high-dimensional case. These two results are all the more remarkable since the Oracle (and thus NLS) was shown to be optimal also for portfolio optimization (theorem 4.1 in Ref.\ \cite{ledoit2017nonlinear}). Let us remark here that reality differs from the assumptions of this theorem in three key ways: the realized covariance matrix is only relevant for a finite number of time steps in the future, this matrix evolves as a function of time, and the calibration data matrix is in the finite-size regime. This implies that in real-life conditions, minimizing the Frobenius norm (or maximizing the Sharpe ratio with calibration data only) is not optimal for portfolio optimization \cite{bongiorno2021oracle}.

\begin{figure}
    \centering

    \includegraphics[width=8.6cm]{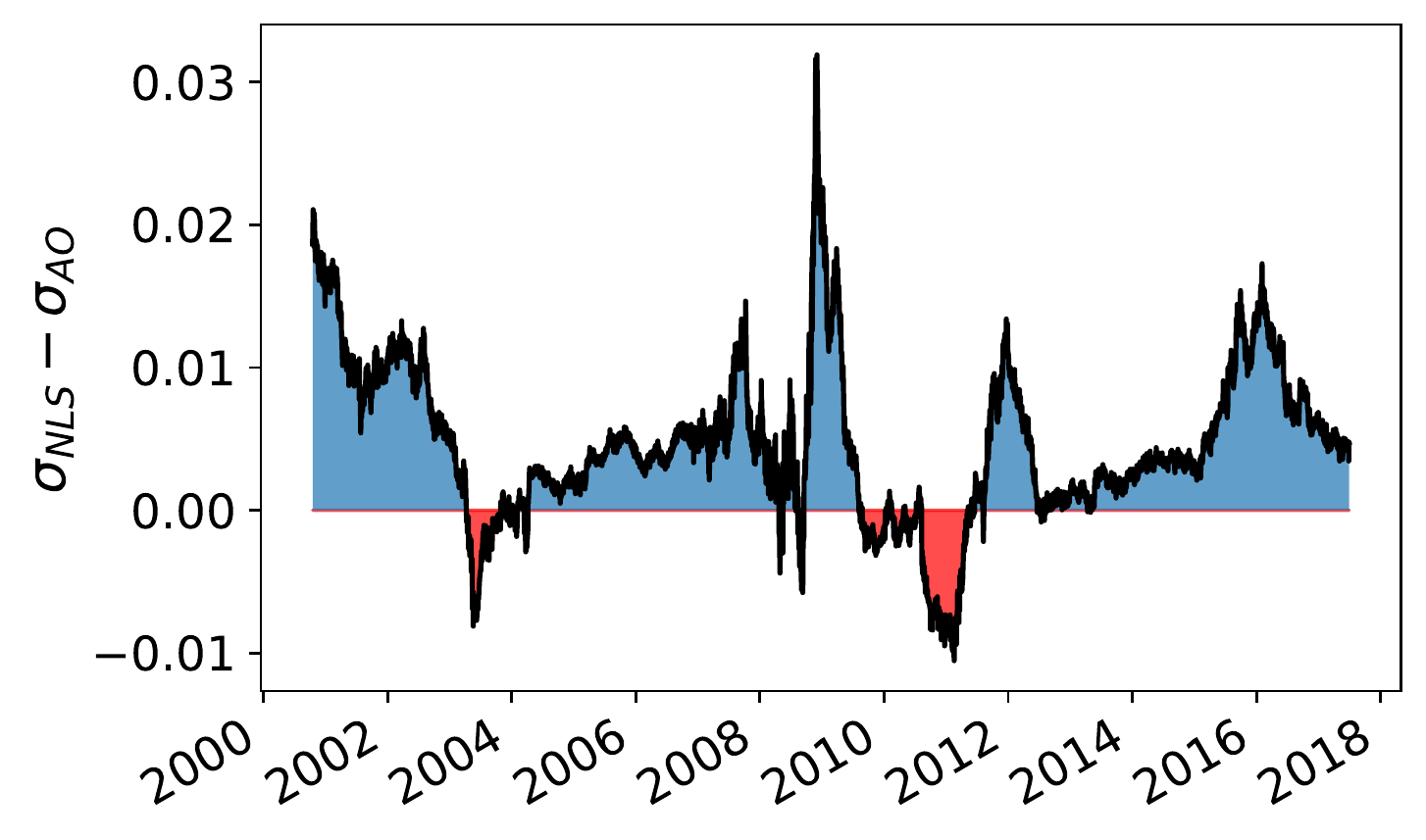}
    \caption{Difference of realized volatility as a function of time between NLS and the Average Oracle. $n=100$, averages over 100 portfolios per date.}
    \label{fig:realized_vol}
\end{figure}

The Average Oracle outperforms  NLS most of the time: plotting the average realized volatility as a function of time (Fig.\ \ref{fig:realized_vol}), one sees that there are only a few periods during which AO loses to NLS, that there seems to be no difference between the AO calibration period (until 2005) and the testing period (from 2006), and finally that the advantage of AO does not decrease as time goes on. 

As a final test, we applied the Average Oracle calibrated with US data to Hong-Kong equity data and found qualitatively similar data (see appendix \ref{sec:HK}). This strongly suggests that the AO captures a systematic time-varying effect found in two different systems with time-varying correlations.

Interestingly, the Average Oracle and the method (CV) we use to compute the NLS \cite{bartz2016cross} share a common ingredient: CV is a $k$-fold cross-validation scheme that computes Oracle eigenvalues from splits of $\mathcal{I}_{\textrm{train}}$ into $k$ sub-intervals and averages them. Equivalently to our case, the two intervals $\mathcal{I}_{\textrm{prev}}$ and $\mathcal{I}_{\textrm{next}}$ will refer to the train and test split of the $k$-fold approach. It is important to point out that restricting the Oracle procedure to the $\mathcal{I}_{\textrm{train}}$ interval implies that at least in $k-1$ of the $\mathcal{I}_{\textrm{next}}$ intervals will have some data points in the past with respect to $\mathcal{I}_{\textrm{prev}}$, leading to an overestimation of the overlap matrix $H^{\circ 2}$. Differently, our method uses 20 years of data and respects causality (time ordering) between $\mathcal{I}_{\textrm{prev}}$ and $\mathcal{I}_{\textrm{next}}$  time windows. The reason why AO outperforms NLS is causality; hence, time ordering needs to be preserved in systems with time-evolving correlations. Note finally that NLS can be computed with other numerical methods which yield similar performance.


We note that NLS can also be applied to z-scores of asset price returns instead of on the returns themselves. In this case, we find that NLS leads to smaller GMV portfolio variance on average, still larger than AO, but with a significantly larger Frobenius distance. This means that the Average Oracle provides a hands-off approach to covariance cleaning and does not require complex computations. Once calibrated, the AO is much faster than NLS.

\section*{Discussion}

By outperforming the optimal estimators for time-invariant covariance matrices, our method shows the need for (and the possibility of) further improvement. Indeed, real systems are rarely time-invariant, and thus one needs to account for the evolution of the real covariance matrix. In such circumstances, we have shown that a simple zeroth-order eigenvalues filtering method that only retains an average dependence between past and future already outperforms known optimal solutions for constant covariance matrices.

The Average Oracle is a first step towards accounting for temporal evolution when filtering covariance matrices, as it is a zeroth-order correction. Beyond time-invariant eigenvalues, AO can also include exponentially weighted moving averages, as in   \cite{potters2005financial,tan2020large}, which results in a slight improvement, as shown by preliminary results. Any additional knowledge about the underlying system may help design higher-order corrections to our method. This knowledge may not come from the covariance matrix itself but from possibly higher-order dependence measures, such as triads, which are much better at predicting the instability of the sign of correlation coefficients (see Ref.\ \cite{bongiorno2020nonparametric}).

The fact that the average influence of the future on the correlation matrix eigenvalues is more informative than the empirical eigenvalues themselves suggests ways to extend analytical results. Similarly, it is natural to quantify how much exploitable information lies in the difference between the empirical and Oracle eigenvalues, i.e., how to mix the Average Oracle eigenvalues with the empirical ones from the training period.

Financial literature proposes complex dynamical models of correlation~\cite{engle2019largecov,pakel2021fitting,denard2021factor,moura2020comparing} that also contains a mechanism to account for the evolution of both eigenvalues and eigenvectors. They are generally computationally much more demanding than AO and rely on strong modeling assumptions. A full comparison between portfolios produced by the naive, model-free, and fast AO and sophisticated DCC-based methods will be reported in future work.


\bibliographystyle{unsrt}
\bibliography{corr_cleaning}

\section*{Acknowledgements}

This publication stems from a partnership between CentraleSup\'elec and BNP Paribas and used HPC resources from the ``M\'esocentre'' computing center of CentraleSup\'elec and \'Ecole Normale Sup\'erieure Paris-Saclay supported by CNRS and R\'egion \^{I}le-de-France.

\textbf{Code:} a notebook is available at  \\ \url{https://gitlab-research.centralesupelec.fr/2019bongiornc/average-oracle-cleaning}.

\appendix

\section{Data}\label{sec:data}
We use about 25 years of daily data for about $N=1000$ US equities from which we compute returns $r_{i,t}= p_i(t)/p_i(t-1)-1$ adjusted for splits, reverse splits and dividends. When computing Oracle eigenvalues, we applied two asset selection filters. 

We applied two asset selection filters. First, for a given $\mathcal{I}^{(b)}_{prev}$ time subinterval and its corresponding $\mathcal{I}^{(b)}_{next}$ subinterval, we only keep the assets that have less than $20\%$ of zero or missing values in the train window in order to avoid undefined standard deviation during the shuffling procedure. Some assets do not have data for the whole period. Other causes for missing data or zero values are technical issues or trading stops. We did not apply the same filter to the test window so as to avoid using future information in our analysis.

In addition, we require that no pair of assets in our subset have a correlation coefficient larger than $0.95$ in the train window to avoid duplicated assets (for example, different asset classes of the same company).

\section{Dealing with large fluctuations of univariate variance}\label{sec:variance}

When the variance of individual time series is not constant, one should apply the AO method on the correlation matrix, i.e., compute the AO eigenvalues on data standardized in  $\mathcal{I}^{(b)}_{\textrm{prev}}$ and $\mathcal{I}^{(b)}_{\textrm{next}}$ separately, and use them to replace the eigenvalues of the correlation matrix computed in $\mathcal{I}_{\textrm{train}}$. As this is clearly the case with financial data, all the results reported here that use financial data use this kind of standardization. The filtered covariance matrix is then defined as the filtered correlation matrix suitably multiplied by the individual standard deviation (see Eq.~\eqref{eq:cov_corr}).

\section{Dependence of the Average Oracle eigenvalues on the next subinterval length}\label{sec:delta_test}

\begin{figure}
    \centering
    \includegraphics[width=8.0cm]{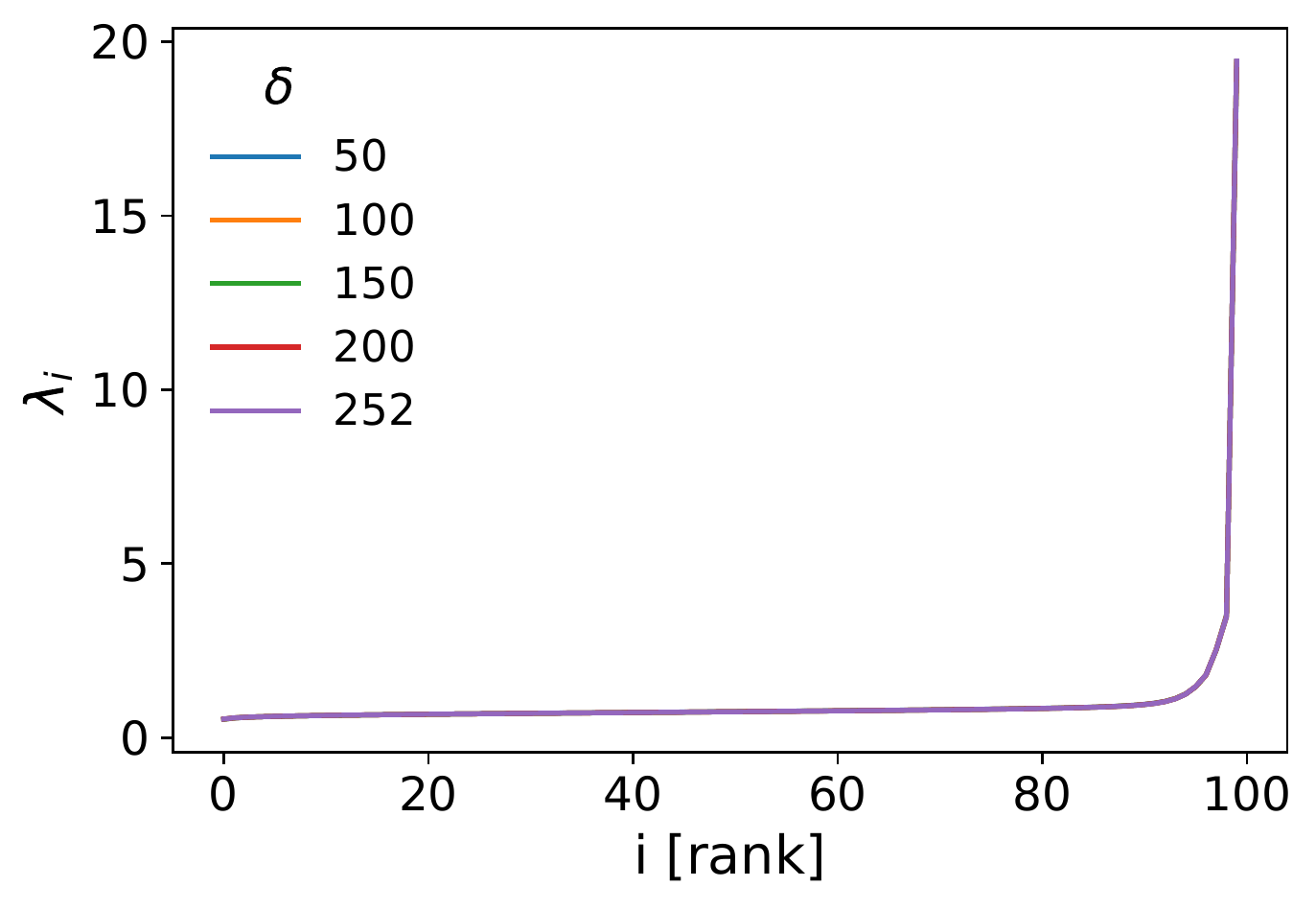}
    \includegraphics[width=8.0cm]{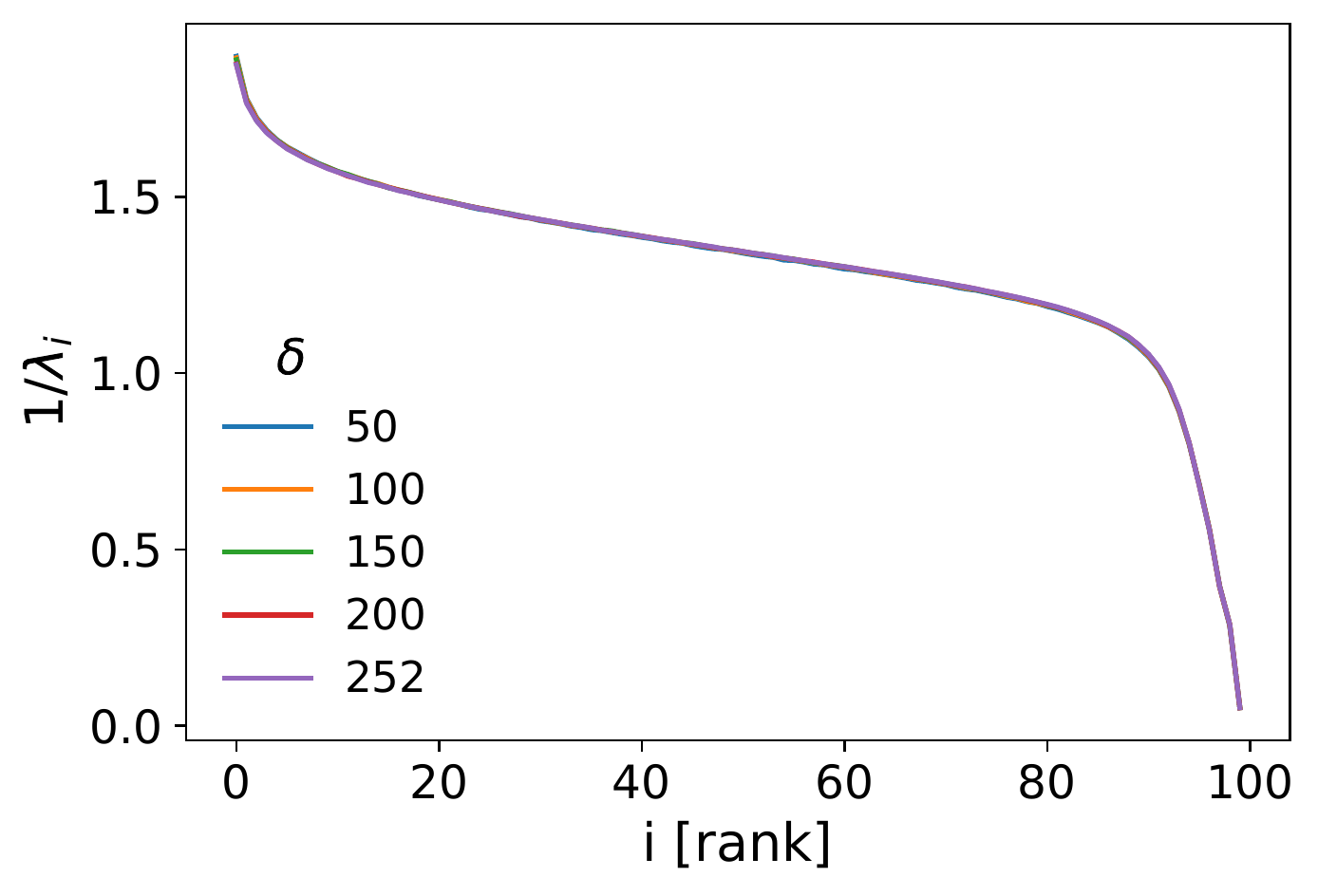}
 \caption{Average Oracle Eigenvalues estimated with a train window  $\delta_{train}=252$ days and various test windows $\delta$. The upper plot displays the averaged eigenvalues, whereas the lower plot refers to their inverse. The calibration is performed with an average over $B=100,000$ elements. }
    \label{fig:cal}
 \end{figure}

We checked that the Oracle eigenvalues can be considered independent from the $\mathcal{I}_{\textrm{next}}$ window length $\delta$. In Fig.~\ref{fig:cal}, we show that the AO eigenvalues for different $\delta$ fixing $\delta_{\textrm{train}}=252$ days:  the estimator is only weakly sensitive to the test window length. This observation supports the idea that the AO procedure can extract the underlying time-invariant part of the eigenvalue dynamic. Note that by reducing the test window, the estimation becomes noisier and thus requires more train and test windows (denoted by $B$) to yield average eigenvalues with the same level of precision.

\section{Influence of overlaps in synthetic data with time-evolving covariance matrices}
\label{sec:model_nonstat}

The empirical findings about overlaps in the main text suggest a simple model with a fixed set of true eigenvalues $\Lambda_{\textrm{true}}$ and a dynamical set of eigenvectors $V_t$. To mimic the real heterogeneity of the eigenvalues, we choose a geometric progression of the eigenvalues, whereas the initial $t=0$ eigenvector basis $V_0$ is chosen randomly from the $SO(n)$ group. At each time-step, the eigenvectors are rotated with rotation matrix $H_t \in SO(n)$, which yields
\begin{equation}
    V_{t+1} = V_{t} H_t.
\end{equation}
In order to control the amount of rotation, we decompose the rotation matrix as $n(n-1)/2$ elementary plane rotation according to the canonical basis, which corresponds to the decomposition of the rotation matrix in Euler angles
\begin{equation}
    H_t = H(\alpha_{1,t}) H(\alpha_{2_t}) \cdots H(\alpha_{n(n-1)/2,t})
\end{equation}
The limit $\alpha_{i,t} \to 0$ corresponds to constant  eigenvectors  and constant covariance. In order to test different levels of time dependence, we sampled independently $\alpha_{i,t} \sim \mathcal{N}(0,s^2)$ from a normal distribution with expected value $0$ and standard deviation $s$.

To simulate the data, we used a factor model
\begin{equation}
    X_{t} = P_{t} A_t,
\end{equation}
where
\begin{equation}
    P_{t} = V_{t} \Lambda^{\frac{1}{2}}_{\textrm{true}}
\end{equation}
and $A_t$ are sampled from $n$ independent standardized normal or  Student t-distributions.

According to this definition, 
\begin{equation}
    \mathcal{E}[X_t X_t^\dagger] =\mathcal{E}[P_t A_t A_t^\dagger P_t^\dagger] = P_t P_t^\dagger = V_t \Lambda_{\textrm{true}} V_t^\dagger 
\end{equation}
since $\mathcal{E}[A_t A_t^\dagger]$ is the identity and $P_t$ are deterministic.

\subsection{Simulations}

Because the eigenvalues are kept fixed in this model, Eq.~(8) stipulates that it is enough to compute the average $\langle H^{\circ 2} \rangle$ in order to obtain the Average Oracle. To this end, we generate 1,000 simulations for each parameter $s$ in the following way. For each simulation, we produce a data matrix $X$ of $n=10$ elements and $T=10,000$ records which represents the full historical dataset. The last $\delta_{test}=50$ records will be kept as the test window, while the first $T-\delta_{test}$ will be the calibration window. The small $n$ choice was necessary due to the non-negligible computational effort to apply the Euler angles rotations.

We first compute the average overlap matrix $\langle H^{\circ 2} \rangle$ from $10,000$ random consecutive time intervals of length $\delta t_{train}=50$ drawn from the calibration window; this is done in two ways, as described in the main text: first by keeping the original time order of the data, and then by building a shuffled data set, which yields the time-dependent and time-independent average overlap matrices. Finally, we compute the train eigenvectors $V_{train}$ from the last $\delta_{train}$ records of the calibration window.

Then, the two AO estimators corresponding to are obtained with 
\begin{equation}
    \lambda_{AO} = H^{\circ 2} \lambda_{\textrm{true}}
\end{equation}
and the RIE estimator as
\begin{equation}
    \Sigma_{AO} = V_{\textrm{train}} \Lambda_{\textrm{AO}} V_{\textrm{train}}^\dagger
\end{equation}
We included the true eigenvalues $\Lambda_{\textrm{true}}$ in the estimator to reduce the amount of noise in the benchmark and to focus on the effect of the eigenvector rotation. 
Finally, we compute the Frobenius distance between the estimators and  $\Sigma_{test}$ from the test window.

 \begin{figure}
    \centering
    \includegraphics[width=8.0cm]{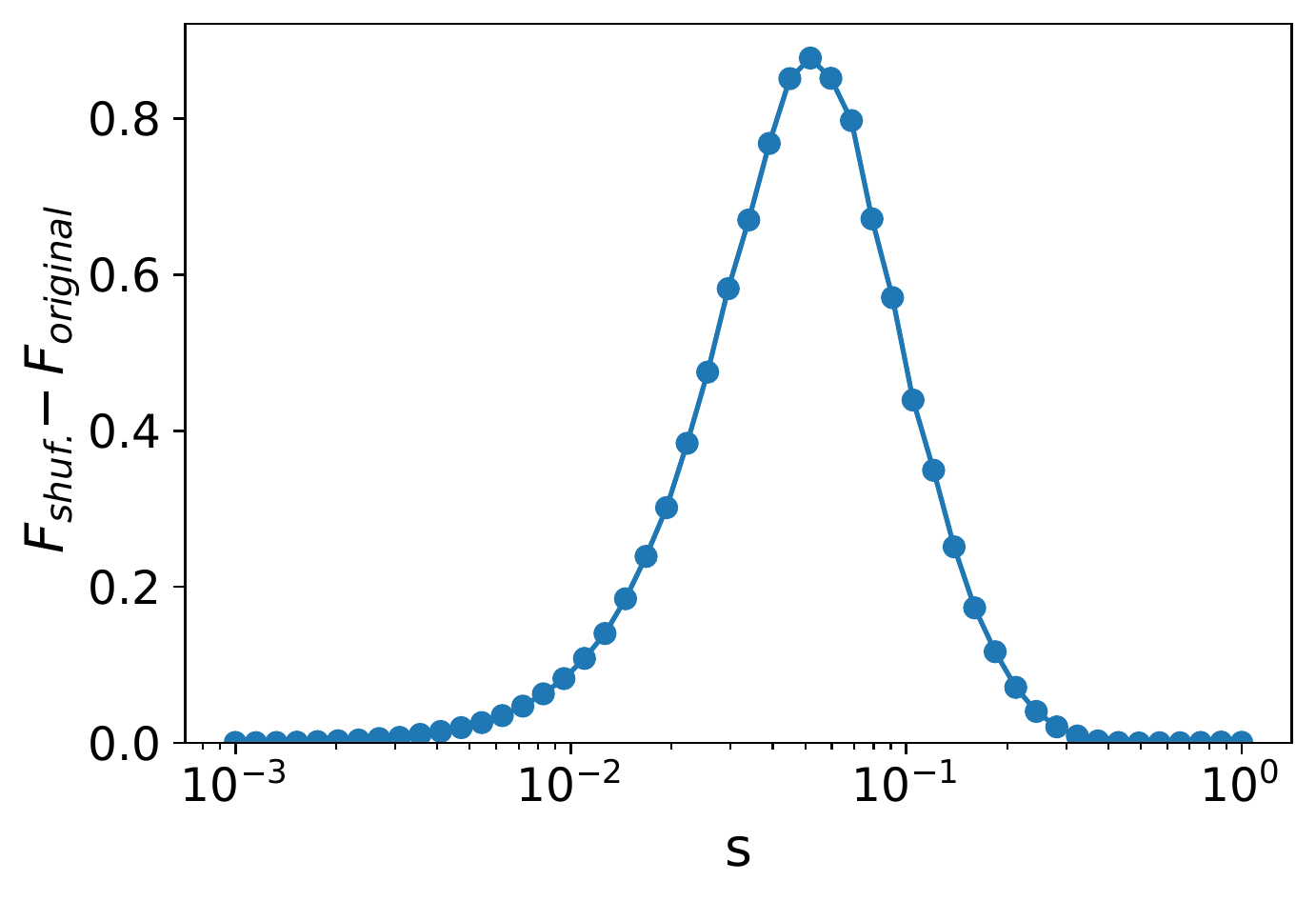}
    \includegraphics[width=8.0cm]{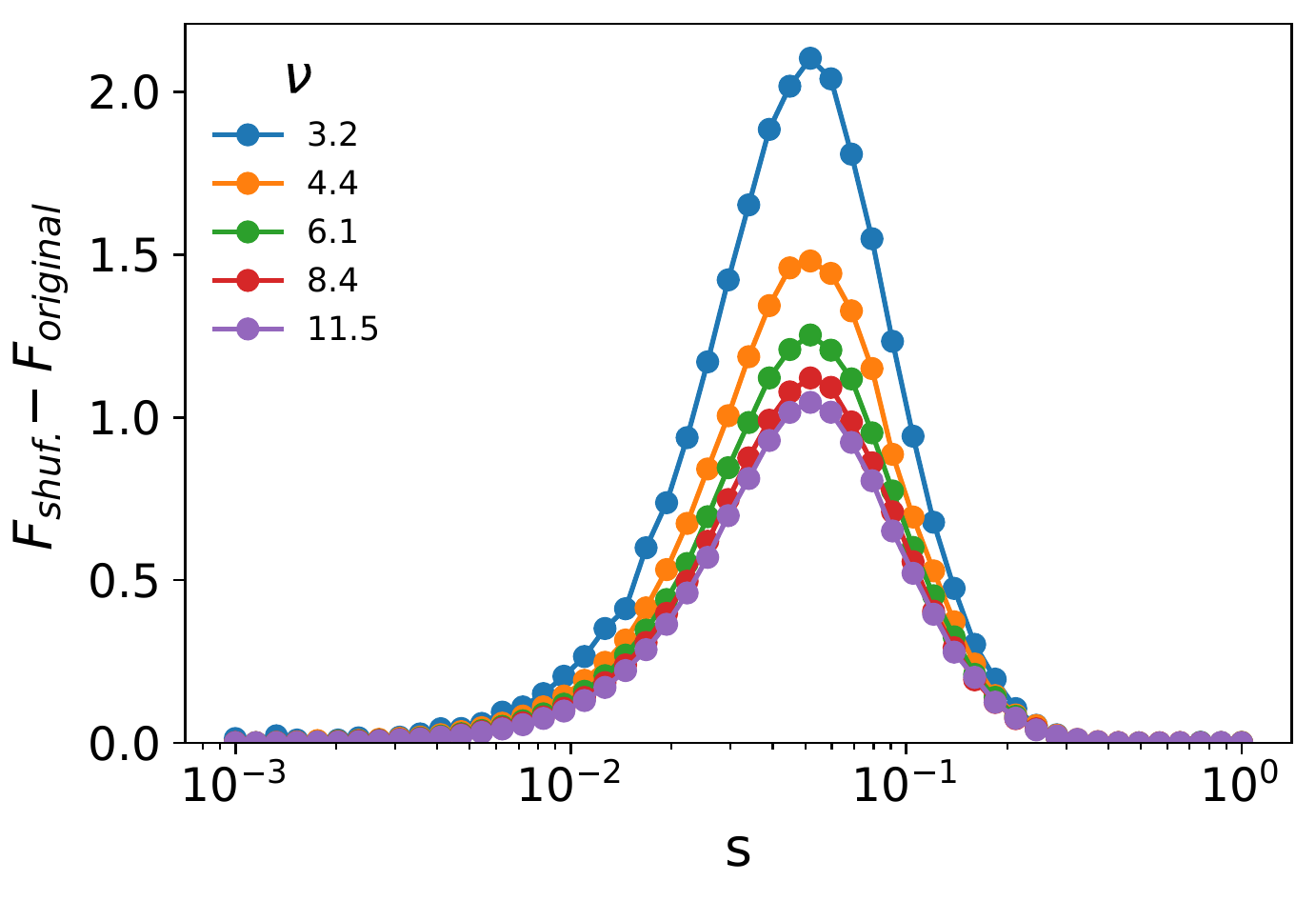}
     \caption{Time evolving benchmark. The upper plot refers to normal variable, and the lower plot to t-student with $\nu$ degrees of freedom. The plots show the difference of Frobenius distance between the test covariance of the time-dependent minus the time-invariant estimator. The x-axis is the average amount of eigenvector rotation over two consecutive time-step. }
    \label{fig:bench2}
 \end{figure}  
In Fig.~\ref{fig:bench2}, we show that for $s\to0$ no difference between the two estimators is detected, as such a case corresponds to the limit of a constant world. Similarly, for $s$ very large, we do not observe any significant difference between the two estimators,  as the temporal evolution is so fast that it destroys any relationship between past and future, and both estimators converge to the identity. Finally, for intermediate values of $s$, the time-dependent estimator outperforms the time-independent one. It is worth reporting that for the case of Student t-distributed variables, the discrepancy between the two estimators increases when the degree of freedom $\nu$ of the t-distribution decreases. This is particularly relevant for financial applications, where the returns are characterized by heavy tails with $\nu\simeq 3$ on average \cite{plerou1999scaling}.

\section{Kullback–Leibler Divergence}\label{sec:KL}
For the sake of completeness, we include a comparison between the covariance estimators and realized covariance matrix by using the Kullback–Leibler (KL) divergence and compare our results with those for the Frobenius distance.

The KL divergence $KL(A||B)$ of the two distributions $A$ and $B$ measures the amount of information lost if $B$ is used to approximate $A$. It is defined as
\begin{equation}\label{eq:KLgen}
    KL(A||B) = \int_{\mathcal{X}} \log \left(\frac{A}{B}\right) dA
\end{equation}
which is the expectation according to distribution $A$  of the log difference $\log(A)-\log(B)$. 
Some authors proposed the KL divergence as an alternative metric to the Frobenius distance when comparing correlation or covariance matrices~\cite{tumminello2007kullback}.

In case of two multivariate central normal distribution data with respective covariance matrices $\Sigma_{test}$ and $\Sigma_{\bullet}$, eq.~\eqref{eq:KLgen} reduces to
\begin{equation}\label{eq:KLnorm}
    KL(\textrm{test}||\bullet) = \frac{1}{2}\left(\textrm{tr}\left[\Sigma_{\bullet}^{-1}\Sigma_{\textrm{test}} \right] -n + \log   \frac{| \Sigma_{\textrm{test}}  | }{| \Sigma_{\bullet }  |}  \right)
\end{equation}
Unfortunately,  there is no closed analytical expression for the KL divergence for multivariate t-distributions 

\begin{figure}
    \centering
    \includegraphics[width=8.6cm]{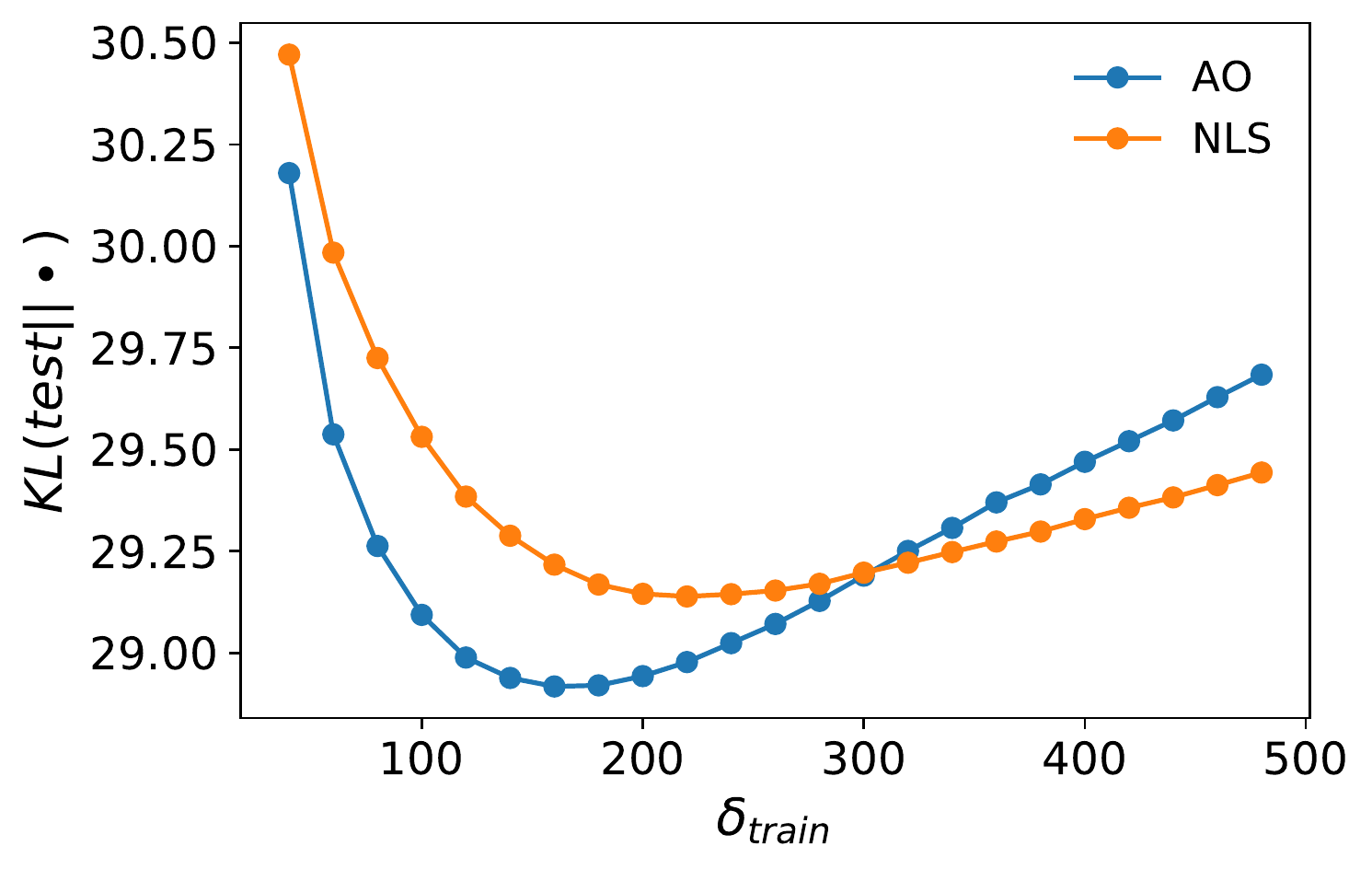}
    \includegraphics[width=8.6cm]{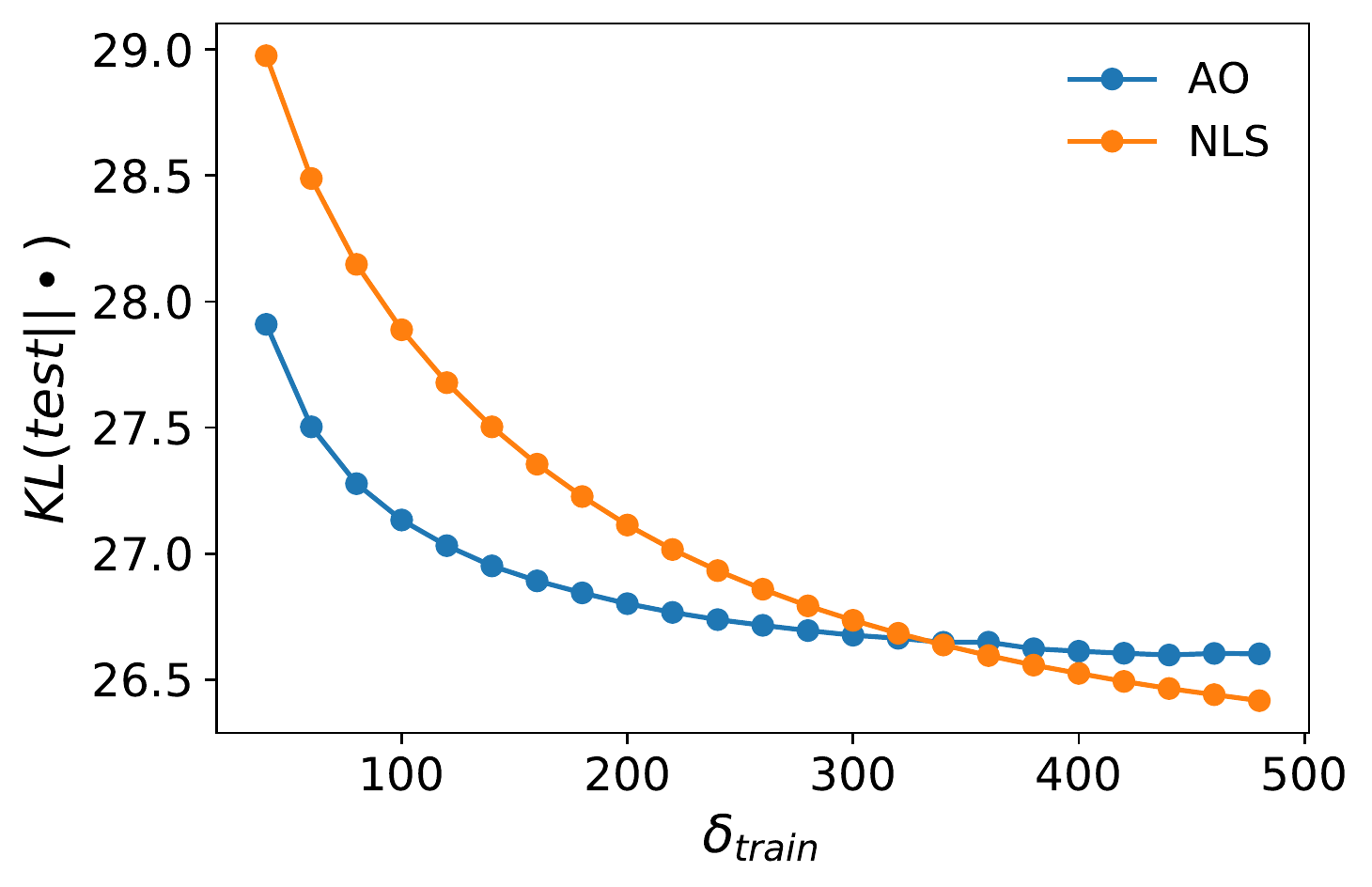}
     \caption{KL divergence between the estimators and test covariance versus the calibration window size $\delta_{train}$. The upper plots refer to the original data set; the lower plot to the shuffled data. The metric is an average over 100 random samples of $n=100$ assets for every day of the out-of-sample period; US equities. The KL divergence is computed in base $n$.  }
    \label{fig:KL}
 \end{figure}

The average KL divergence in base $\log(n)$ over the validation window $[2006,2018]$ is reported in Fig.~\ref{fig:KL}. It is important to remark that Eq.~\eqref{eq:KLnorm} is not defined if one of the two covariance matrices has at least one null eigenvalue. While the RIE is always positive-defined, it is possible that the test covariance matrix is not. Those cases were not considered in our analysis. 

The results for the KL divergence reported for the time-ordered data are compatible with those for the Frobenius norm: the AO estimator outperforms NLS, with an optimal calibration window around $\delta_{train}=180$ days. For the shuffled case, we observe an important difference from the Frobenius norm results since AO outperforms NLS for small training windows $\delta_{train}<300$ days. The behavior of both curves, in this case, is monotonically decreasing, which is expected since, in a  world with constant covariance matrices, more data always yields a better estimation. We suspect that such difference might be due to a higher sensitivity of the KL divergence to overfitting in the train window; thus, a small $\delta_{train}$ should affect more NLS than AO. The fact that NLS outperforms AO for large $\delta_{train}$ in the shuffled case is to be expected, as NLS is optimal in the asymptotic regime of $n$ and $\delta_{train}$ large for time-independent covariance matrices.

\section{Optimal Test Window Length of the Estimators}\label{sec:test_windowlength}
In this section, we explored how the performance changes as the length of the test window $\delta_{test}$ varies. 

 \begin{figure}
    \centering
    \includegraphics[width=8.0cm]{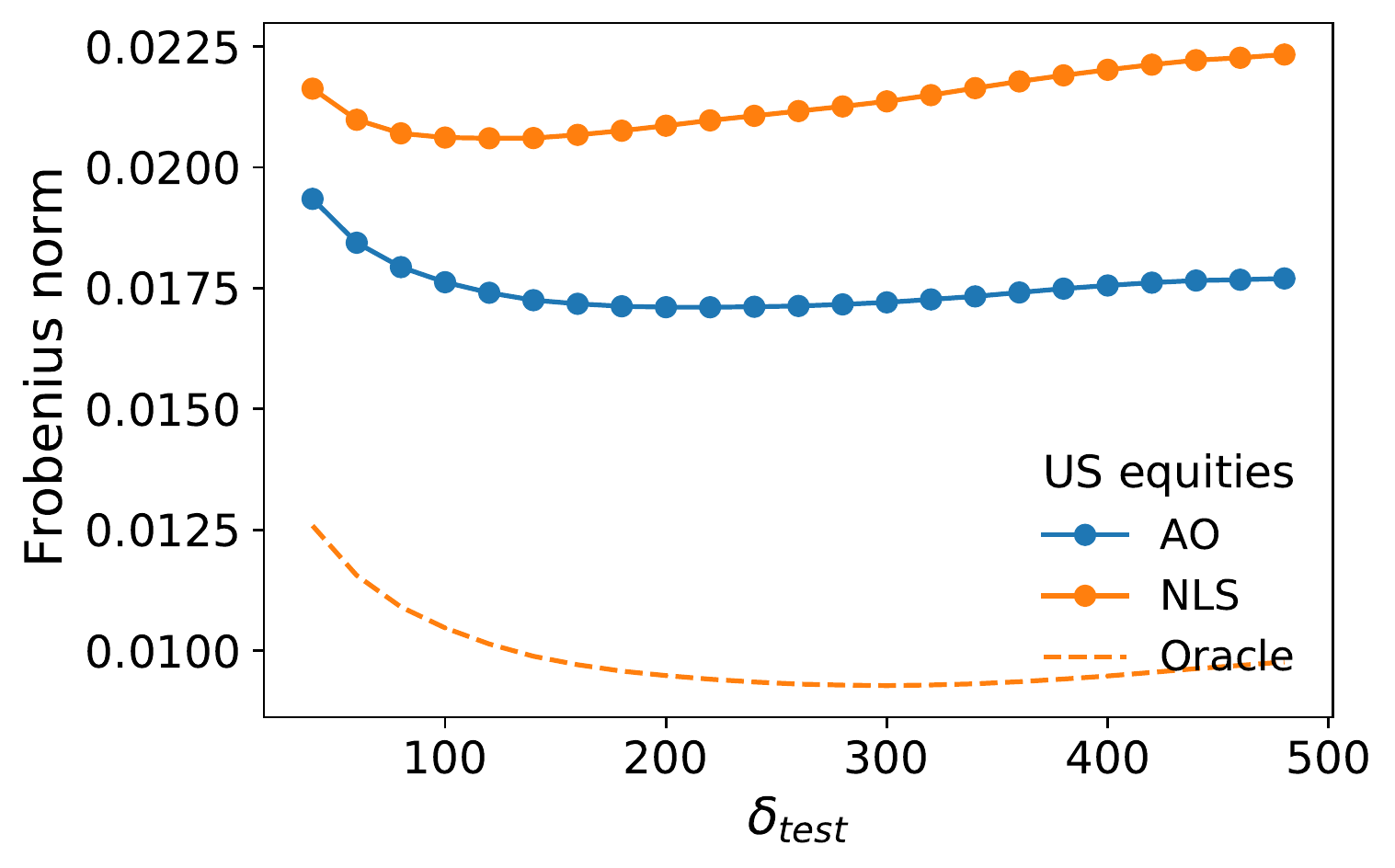}
    \includegraphics[width=8.0cm]{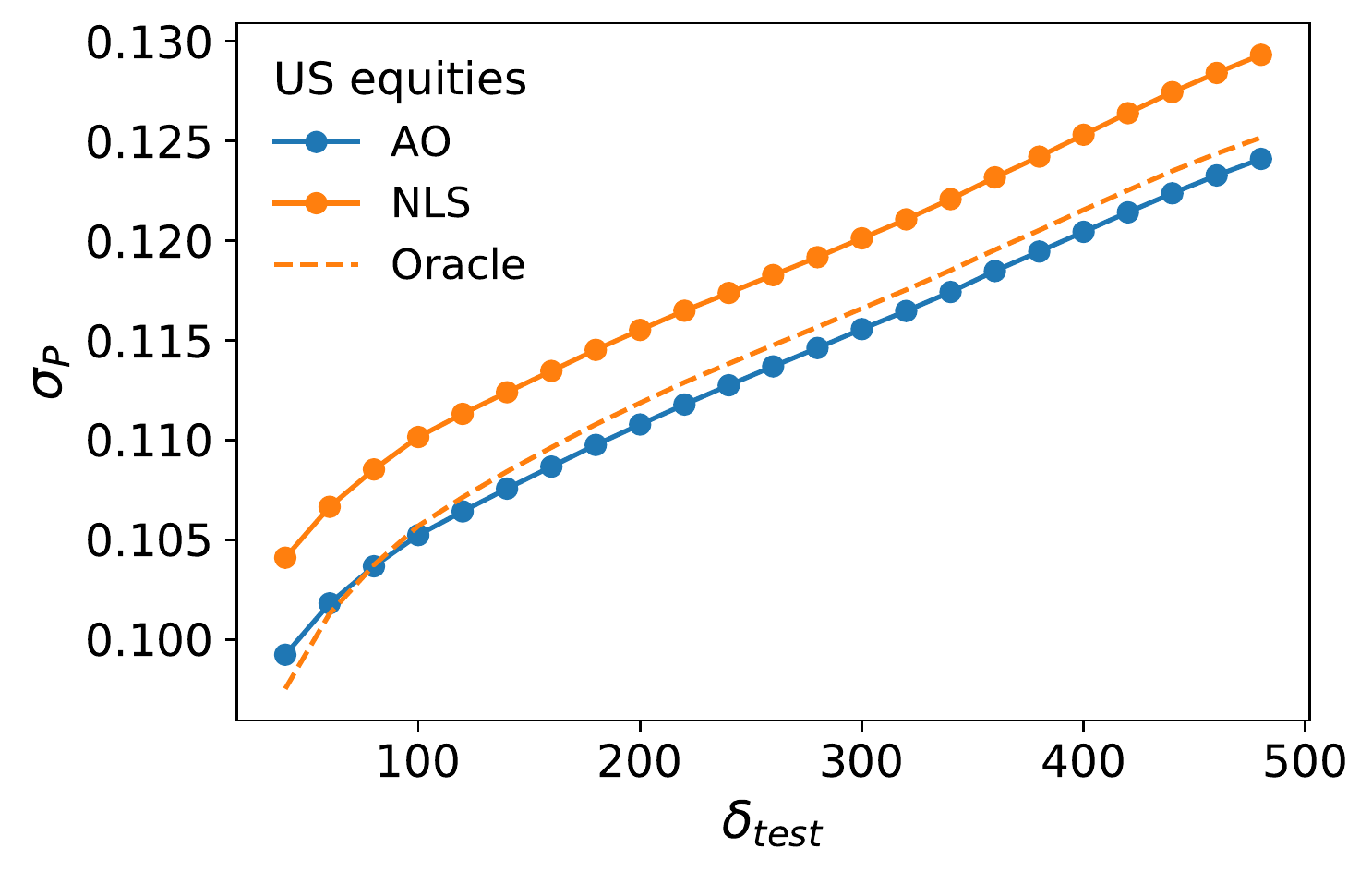}
     \caption{Out-of-sample performance measures of the filtered covariance matrices as a function of the test window length. The upper plot refers to the average Frobenius distance; the lower plot refers to the realized volatility of the GMV portfolio. We fixed $\delta_{train}=200$ days, and $n=100$ random assets are selected $100$ times for each day of the out-of-sample period.}
    \label{fig:tout}
 \end{figure}  
 
In the upper panel of Fig.~\ref{fig:tout} we show the Frobenius norm between the estimators and the test covariance. For very small time horizons, all estimators reach a high distance with the test covariance. This is probably due to the high sample size error on the test covariance if it is computed over a short time window. In addition, we observe that the Frobenius norm of the NLS estimator has a minimum at around $\delta_{test}=100$ days, a further increase of $\delta_{test}$ decreasing its performance. On the other hand, the AO estimator is much more stable for large $\delta_{test}$. This supports the intuition that AO can really extract the time-invariant part of the system evolution. 

Looking at the volatility of the GMV portfolio, we observe that the global minimum is reached for all estimators at the smallest $\delta_{test}$ then it increases approximately linearly with $\delta_{test}$. The latter is not in contradiction with the results of the Frobenius norm since the global optimal minimum of the volatility changes as $\delta_{test}$ increases. In simple words, with a single portfolio held for two years, it is impossible to reach the same low volatility of a portfolio updated weekly.

\section{System size dependence}
In this section, we explored how the performance changes as the system size $n$ varies. 

 \begin{figure}
    \centering
    \includegraphics[width=8.0cm]{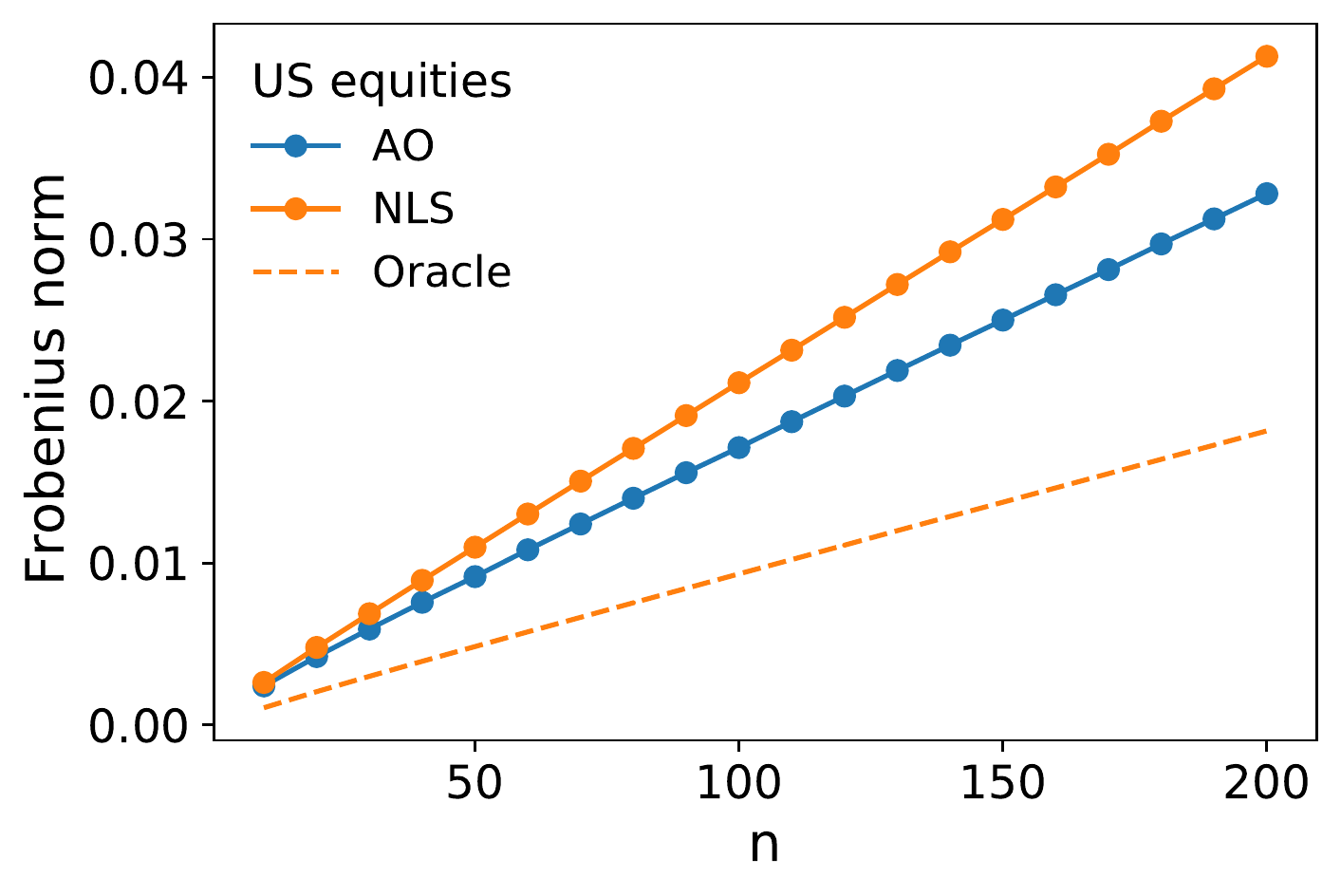}
    \includegraphics[width=8.0cm]{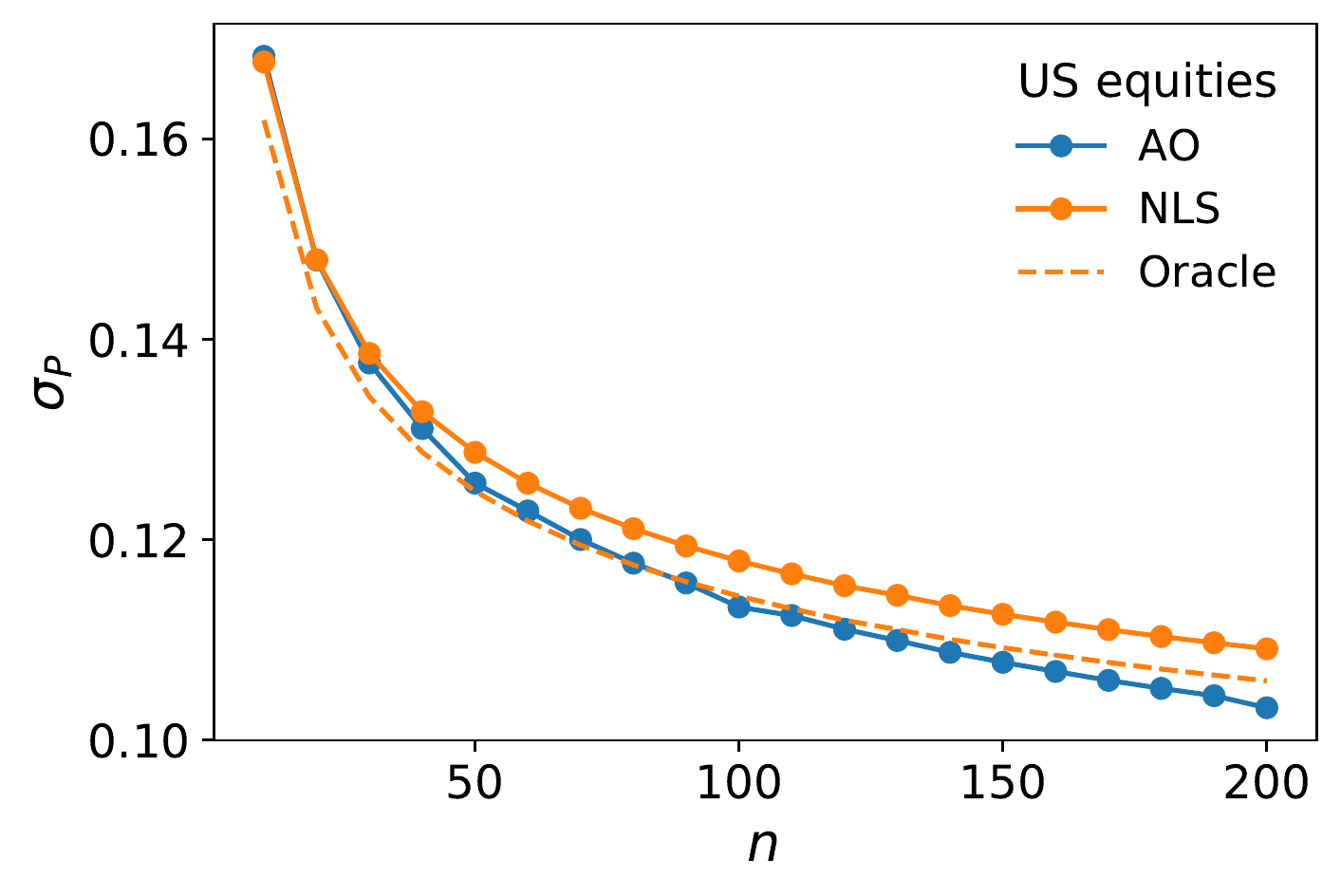}
     \caption{Out-of-sample performance measure of the filtered covariance matrices as a function of the system size. The upper plot refers to the average Frobenius distance; the lower plot to the realized volatility of GMV portfolios. We fixed $\delta_{train}=200$ days and $\delta_{test}=252$ days,$n$ random assets are selected $100$ times for each day of the out-of-sample period.}
    \label{fig:n}
 \end{figure}

In the upper panel of Fig.~\ref{fig:n} we show the Frobenius norm distance for the three estimators. As $n$ increases, the distance between the three estimators and the test covariance matrix increases. This is expected since the eigenvalue correction is only applied to $O(n)$ degrees of freedom, whereas the Frobenius norm increases as $n (n-1) /2$. It is worth remarking that the AO estimator outperforms the NLS for all the values of $n$ 

In the lower panel of Fig.~\ref{fig:n} we show the GMV portfolio volatility as a function of $n$.
As in the previous case, changing the number of stocks changes the optimal minimum reachable on the test window: the larger $n$, the more possibilities one has to obtain a low-volatility portfolio. In addition, we observe that the relative discrepancy between the AO and NLS estimator increases with $n$.

\section{Hong Kong Stock Exchange}\label{sec:HK}
\begin{figure}
    \centering
    \includegraphics[width=8.0cm]{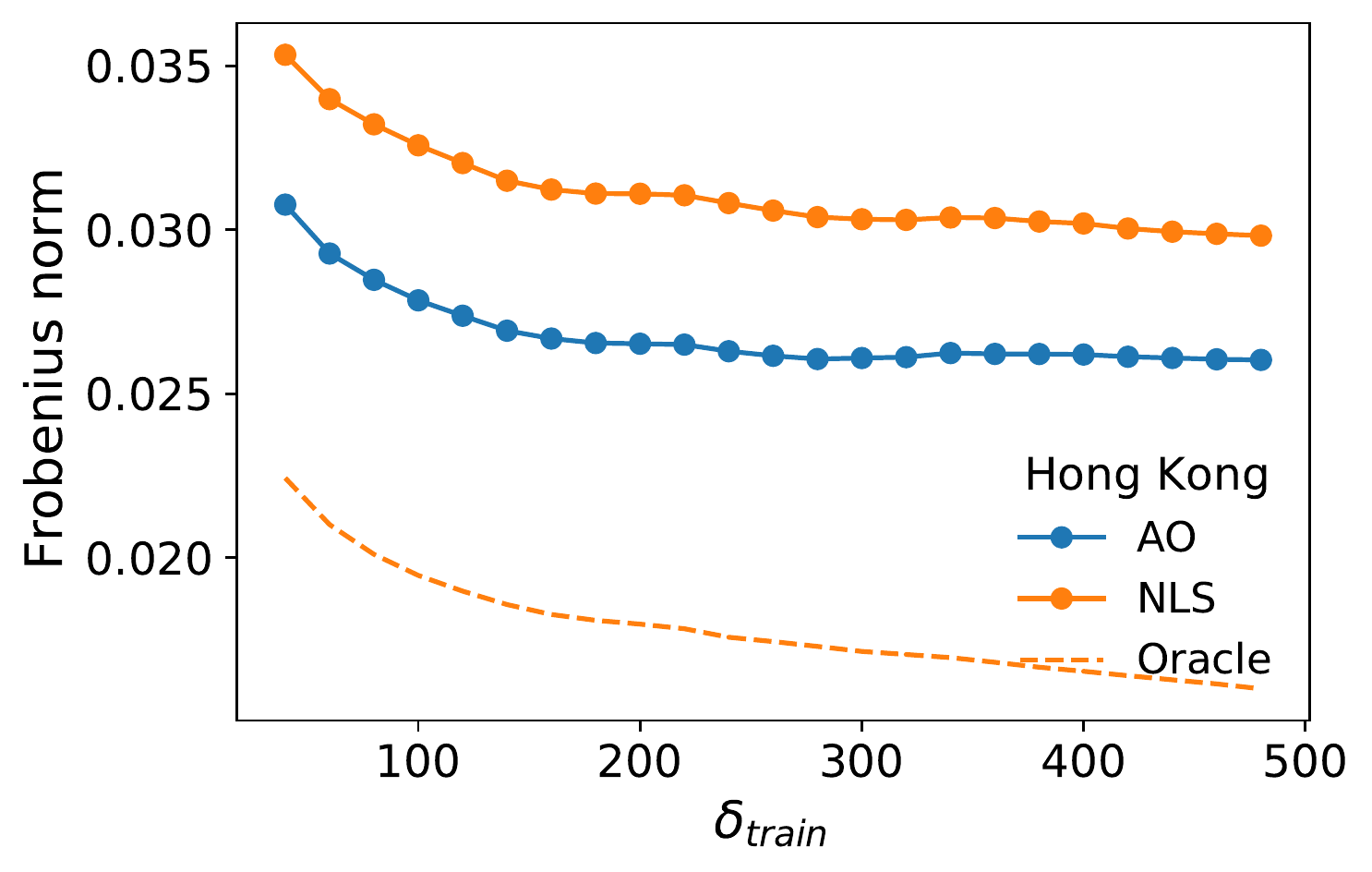}
    \includegraphics[width=8.0cm]{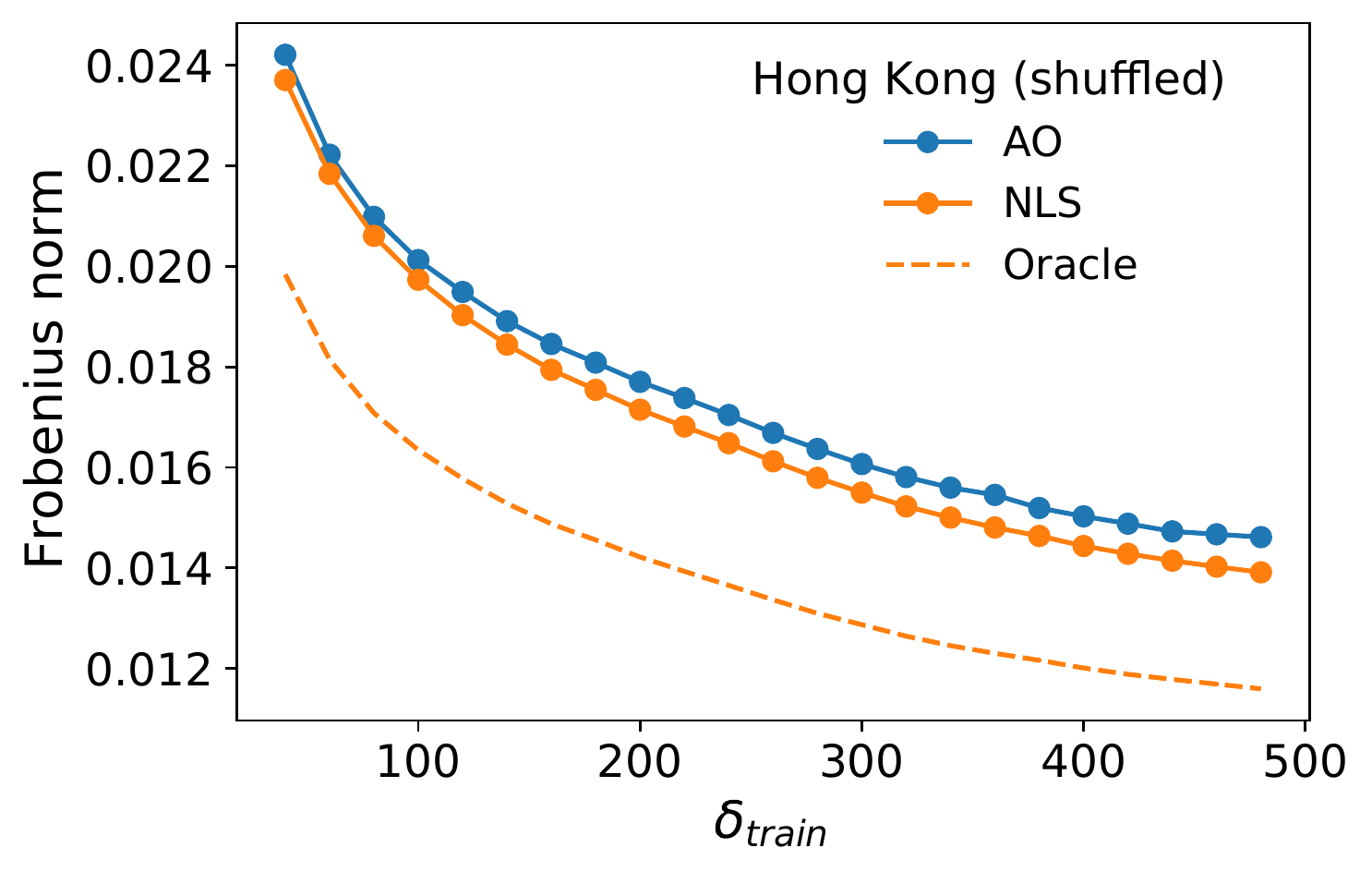}
     \caption{Average Frobenius distance between the filtered and test covariance matrices as a function of the calibration window length in the out-of-sample period. The upper plot refers to the original data set; the lower plot to the shuffled data. 100 portfolios with  $n=100$ random assets are computed for each day of the out-of-sample period.}
    \label{fig:L2_HK}
 \end{figure}

We took the AO eigenvalues calibrated with US equities data and used them to filter covariance matrices from the Hong Kong stock exchange data from the  [2004-01-01,2017-06-23] period. In Fig.~\ref{fig:L2_HK} we show the Frobenius distance between the covariance estimator and the out-of-sample covariance matrix. As for US equities, the AO provides a better estimator of the out-of-sample covariance matrix for the regular time series while being worse for shuffled data.

\begin{figure}
    \centering
  
    \includegraphics[width=8.0cm]{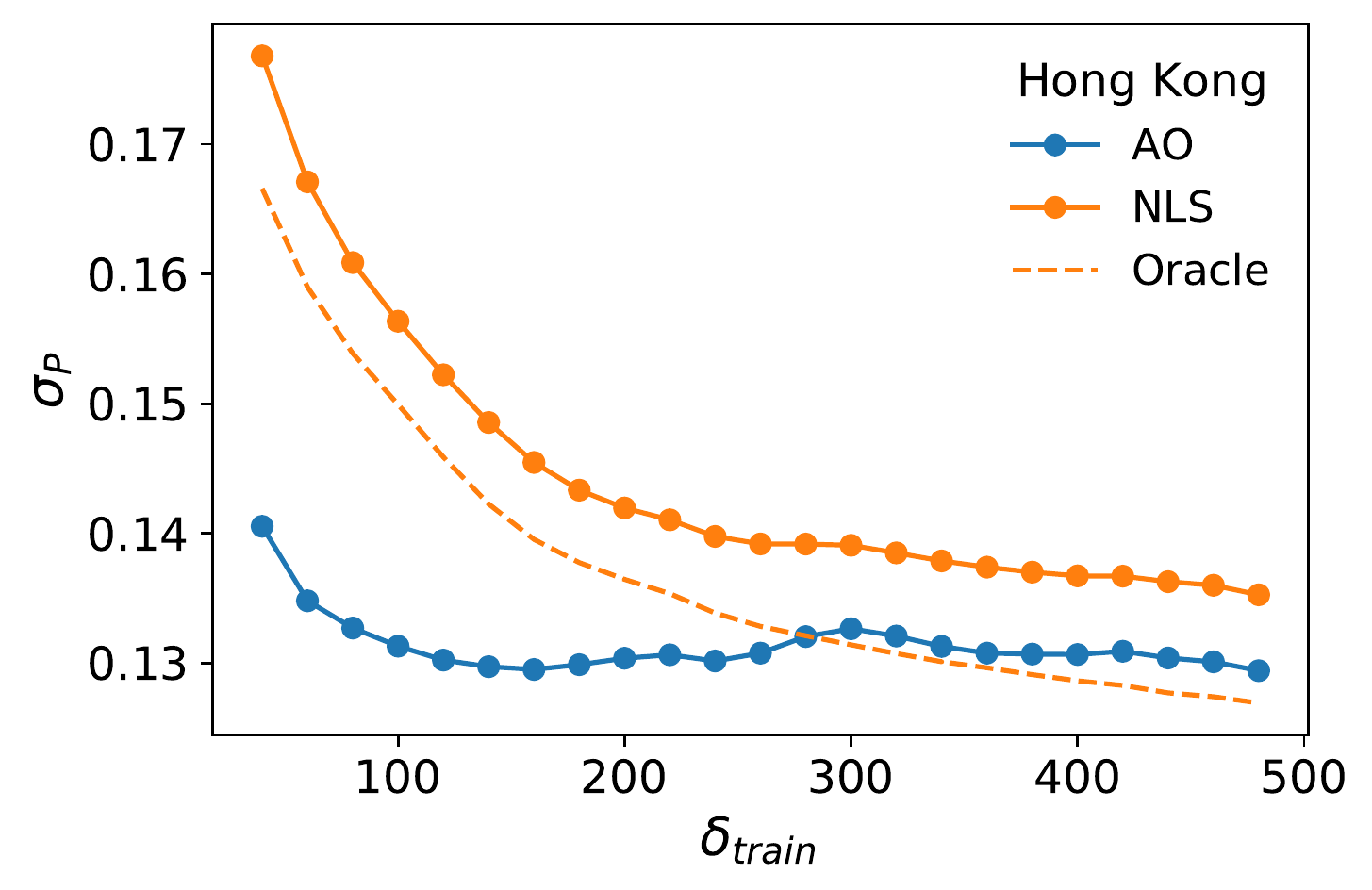}
    \includegraphics[width=8.0cm]{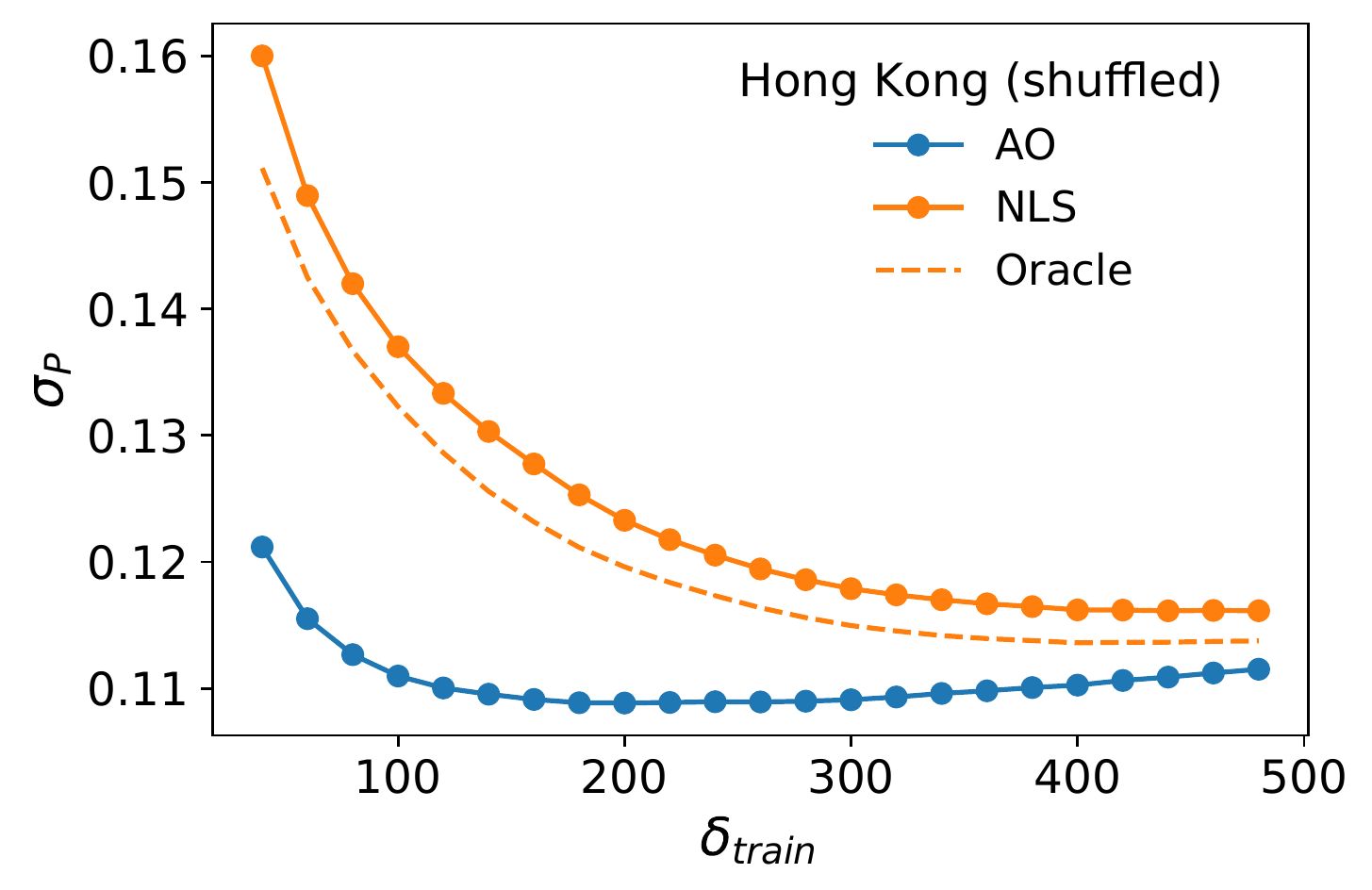}
    \caption{Average realized volatility of Global Minimum Variance portfolios as a function of the calibration window length. The upper plot refers to the original dataset and the lower plot to the shuffled data. 100 portfolios with $n=100$ random assets are computed for each day of the out-of-sample period.}
    \label{fig:vol_HK}
\end{figure}

In Fig.~\ref{fig:vol_HK}, we show the realized variance of global minimum portfolios. As for the US equities, the AO yields lower variance than the Oracle for short calibration windows; AO also beats NLS for all the calibration window lengths that we tested, both for the regular and shuffled cases.  
 
\end{document}